\documentclass[useAMS, usenatbib]{mn2e}

\newcounter{footnew}
\newcommand{\startfoot}{\setcounter{footnew}{0}}
\newcommand{\valfoot}{\stepcounter{footnew}\footnotemark[\value{footnew}]}
\newcommand{\newfoot}{
\valfoot
\renewcommand*{\thefootnote}{\arabic{footnote}}
}
\newcommand{\textfoot}[1]{\newfoot{} {\footnotesize #1 }\\}

\usepackage{natbib}
\usepackage{graphicx}
\usepackage{amssymb}
\usepackage{slantsc}
\usepackage{ifpdf}
\usepackage{rotating}
\usepackage{tablefootnote}
\usepackage{pdflscape}
\usepackage{longtable}
\usepackage{journals}
\usepackage{amsmath}
\usepackage{color}
\usepackage{supertabular,booktabs}
\usepackage[titletoc]{appendix}
\usepackage{adjustbox}
\newcommand{\caps}[1]{{\scshape{#1}}}

\begin{document}

\title[Extreme quiescent variability of EXO 1745--248]{Extreme Quiescent Variability of the Transient Neutron Star Low-mass X-ray Binary EXO 1745--248 in Terzan 5}
\author[L.E. Rivera Sandoval et al.]{L. E. Rivera Sandoval$^{1,2}$\thanks{E-mail:liliana.rivera@ttu.edu} 
R. Wijnands$^1$, N. Degenaar$^1$, Y. Cavecchi$^{3,4}$, C. O. Heinke$^5$,
\newauthor E. M. Cackett$^6$, J. Homan$^{7,8,9}$, D. Altamirano$^{10}$, A. Bahramian$^{11}$, G. R. Sivakoff$^5$, 
\newauthor J. M. Miller$^{12}$ \& A. S. Parikh$^1$\\
\\
$^1$ Anton Pannekoek Institute for Astronomy, University of Amsterdam, Science Park 904, 1098 XH Amsterdam, The Netherlands\\
$^2$ Department of Physics, Box 41051, Science Building, Texas Tech University, Lubbock, TX 79409-1051, USA\\
$^3$ Department of Astrophysical Sciences, Princeton University, Peyton Hall, Princeton, NJ 08544, USA\\
$^4$ Mathematical Sciences and STAG Research Centre, University of Southampton, SO17 1BJ, UK \\
$^5$ Department of Physics, University of Alberta, CCIS 4-183, Edmonton, AB$_{435}$T6G 2E1, Canada\\
$^6$ Department of Physics \& Astronomy, Wayne State University, 666 W. Hancock St, Detroit, MI 48201, USA\\
$^7$ MIT Kavli Institute for Astrophysics and Space Research, 77 Massachusetts Avenue 37-582D, Cambridge, MA 02139, USA \\
$^8$  SRON, Netherlands Institute for Space Research, Sorbonnelaan 2, 3584 CA Utrecht, The Netherlands\\
$^9 $ Eureka Scientific, Inc., 2452 Delmer Street, Oakland, CA 94602, USA\\
$^{10}$  Physics and Astronomy, University of Southampton, Southampton SO17 1BJ, UK\\
$^{11}$ Department of Physics and Astronomy, Michigan State University, East Lansing, MI, USA\\
$^{12}$ Department of Astronomy, University of Michigan, 1085 South University Ave, Ann Arbor, MI 48109-1107, USA\\
}

\pagerange{\pageref{firstpage}--\pageref{lastpage}} \pubyear{2002}

\maketitle

\label{firstpage}

\begin{abstract}

EXO 1745--248 is a transient neutron-star low-mass X-ray binary that resides in the globular cluster Terzan 5. 
We studied the transient during its quiescent state using 18 $Chandra$ observations of the cluster acquired between 2003 and 2016.
We found an extremely variable source, with a luminosity variation in the 0.5--10 keV energy range of $\sim3$ orders of magnitude (between $3\times10^{31}$ erg s$^{-1}$ and $2\times10^{34}$ erg s$^{-1}$) on timescales from years down to only a few days. 
Using an absorbed power-law model to fit its quiescent spectra, we obtained a typical photon index of $\sim1.4$, indicating that the source is even harder than during outburst and much harder than typical quiescent neutron stars if their quiescent X-ray spectra are also described by a single power-law model. This indicates that EXO 1745--248 is very hard throughout the entire observed X-ray luminosity range. 
At the highest luminosity, the spectrum fits better when an additional (soft) component is added to the model. All these quiescent properties are likely related to strong variability in the low-level accretion rate in the system. However, its extreme variable behavior is strikingly different from the one observed for other neutron star transients that are thought to still accrete in quiescence.
We compare our results to these systems. We also discuss similarities and differences between our target and the transitional millisecond pulsar IGR J18245--2452, which also has hard spectra and strong variability during quiescence.

\end{abstract}


\section{Introduction}

A low mass X-ray binary (LMXB) is a system in which a low mass star (typically $\lesssim$1 M$_{\sun}$) has filled its Roche lobe and transfers mass to a 
 neutron star (NS) or a black hole (BH). The type of accretor can often be identified as being a NS when thermonuclear X-ray bursts or X-ray pulsations are detected since these phenomena require the presence of a solid surface and/or a magnetic field.  

Transient LMXBs are systems that undergo episodes of active accretion (called outburst)
after long periods (lasting up to decades) of quiescence. 
These outburst episodes typically last at most several months, though a few LMXB outbursts are known to last for several years or even decades. 
Outbursts are thought to be caused by instabilities in the accretion disc \citep[see the review by][]{2001lazo} and can reach peak X-ray luminosities of $L_X \sim 10^{35-39}$ erg s$^{-1}$. During the quiescence period no or only little accretion of matter occurs and the LMXBs are observed at X-ray luminosities of only $10^{30-34}$ erg s$^{-1}$. In this paper we have used the term ``quiescence" to refer to the state X-ray binaries are in when they have X-ray luminosities $<10^{34}$ erg s$^{-1}$ \citep[a definition commonly used in the literature, e.g.][]{2013plotkin,2015wij}.  Below this luminosity limit, systems that harbour a BH become softer when their X-ray luminosity decreases
\citep[likely due to a change in the properties of the accretion flow, e.g.][]{2013plotkin}. In contrast, NS systems exhibit a more complex and heterogenous behaviour \citep[e.g., depending on source, the quiescent spectrum might be dominated by different spectral components; see][for a in-depth discussion]{2015wij}.

The X-ray spectra of the quiescent systems that harbour BHs \citep[see e.g.][]{2002kong,2003hameury,2003ATom,2013plotkin} and the spectra of a subset of the NS systems \citep[e.g.][]{2002campana, 2005wijnands,2007heinke,2009heinke,2012dege,2012degena} can be described by a hard component, which is typically modeled with a phenomenological power-law component. In the BH systems, this component has to come from some form of accretion onto the BH \citep[see][and references therein for a detailed discussion]{2013plotkin}, but for the NS systems the origin of this hard component is not clear. It could be related to accretion as well, to physical processes that involve the NS magnetic field \citep[see e.g.][]{2012degena,2013Bernardini},
or for example to shock emission between the incoming matter and the relativistic radio pulsar wind \citep{1998campana}.

Although a significant number of quiescent NS systems have hard spectra, the majority of the NS systems either have pure soft, thermal quiescent spectra 
\citep[typically described by a neutron-star atmosphere model, e.g.][]{2003wijnands,2004wijands,2011degenaar,2012lowell,2014Homan} or a two component spectrum in which the thermal and the non-thermal components both contribute significantly \citep[e.g.,][]{1996asai,1998Asai,2001zand,2001rut,2002rut,2004tom,2005cac}. When the spectra are strongly dominated by the soft component, it is typically thought that the emission originates from the cooling emission of a NS tht has been heated due to the accretion of matter \citep[see the review by][]{2017wijnands}. When a strong hard component is present as well, the origin of both components is less well understood. 
Although accretion onto the NS surface had been considered as the reason for the soft component \citep[][]{1998campana}, only recently\footnote{Although earlier studies already suggested that the power-law component could also be due to low-level accretion onto the neutron star \citep[e.g.,][]{2002rut,2010cacket}, the evidence for this was still rather inconclusive.} strong evidence emerged that indeed \textit{both} components arise from accretion onto the NS, with the soft component coming directly from the surface (due to potential energy release when the matter hits the surface) and the hard component originating from the boundary layer \citep[through Bremsstrahlung;][]{2014cha,2015dangelo}. \citet{2015wij} suggested that if both components contribute $\sim50\%$ to the unabsorbed 0.5--10 keV flux, both components likely are due to accretion \citep[see also][]{2010cacket}.

Transient LMXBs have been identified in different environments of our Galaxy.  
Most are located in the disc of the Galaxy, but some of them reside in globular clusters (GCs), such as Terzan 5.
This GC is located at $5.5\pm0.9$ kpc \citep{2007orto} in the direction towards the bulge of the Galaxy. Thanks to the high spatial resolution of the \textit{Chandra X-ray Observatory},
many faint X-ray sources have been detected in this cluster \citep[e.g.][]{2006Heinke} when no transient was in outburst. When in outburst, a bright transient makes studying these faint sources very difficult and, in fact, frequent X-ray outbursts have been observed from sources located in the cluster. \textit{Chandra} observations during these outbursts allowed for accurate positions of the associated transients \citep{2003heinke,2010pool,2011pool,2012homan} and so far three distinct X-ray transients have been identified there (all NS systems; see for example Table 1 of \citeauthor{2012dege}, \citeyear{2012dege} and Table 5 of \citeauthor{2014bara}, \citeyear{2014bara}). The transient NS LMXB EXO 1745--248 has been identified as the system responsible for several of the episodes of high X-ray activity in the cluster \citep[see, e.g.,][]{2003heinke,2012serino,2012dege,2016wijnands,2016teta,2017matra}. Observations of thermonuclear bursts from this source have demonstrated that the primary object is a NS (see e.g. Table 1 in \citeauthor{2012dege}, \citeyear{2012dege} and references therein)

The quiescent state of EXO 1745--248 has been investigated by \citet{2005wijnands}  and \citet {2012dege}
using $Chandra$ observations of Terzan 5 when the transient was not in outburst. Both studies found that the quiescent X-ray spectrum of the source could be modeled with only a hard emission component with no need to add a soft component. In addition, \citet{2012dege} found that the source was very variable between different observations; it exhibited a quiescent luminosity between $4\times10^{31}$ erg s$^{-1}$ and $\sim10^{33}$ erg s$^{-1}$ during the period 2003--2011. In order to determine an upper limit on the contribution of a possible thermal component in the quiescent spectrum, \citet{2012dege} added a NS atmosphere component to their power-law model. They
determined a NS surface temperature of $\lesssim42$ eV. 
This low upper limit indicates that the NS in EXO 1745--248 has a relatively cold core, suggesting that the NS efficiently cools in-between outbursts
\citep[for example through enhanced neutrino emission;][]{2004yaco,2012dege}.

In this paper, we further study the quiescent spectra and the extreme variability of EXO 1745--248. 
By studying the unusual quiescent behavior of this source, we increase our general understanding of quiescent NS LMXB systems. 
To accomplish our goals, we complement the $Chandra$ observations used by \citet{2005wijnands} and \citet{2012dege} with 14 additional observations taken during the period $2011-2016$. 
Several of them have been used previously to study the other transients in Terzan 5 \citep[e.g.][]{2015deg_c,2013dege_c,2014bara}. 
However, it is the first time that these additional observations have been used to further study the quiescent behaviour of EXO 1745-248.

\begin{figure}
\centering
  \includegraphics[width=.5\textwidth, trim=1.cm 0.0cm 1.0cm 1.5cm]{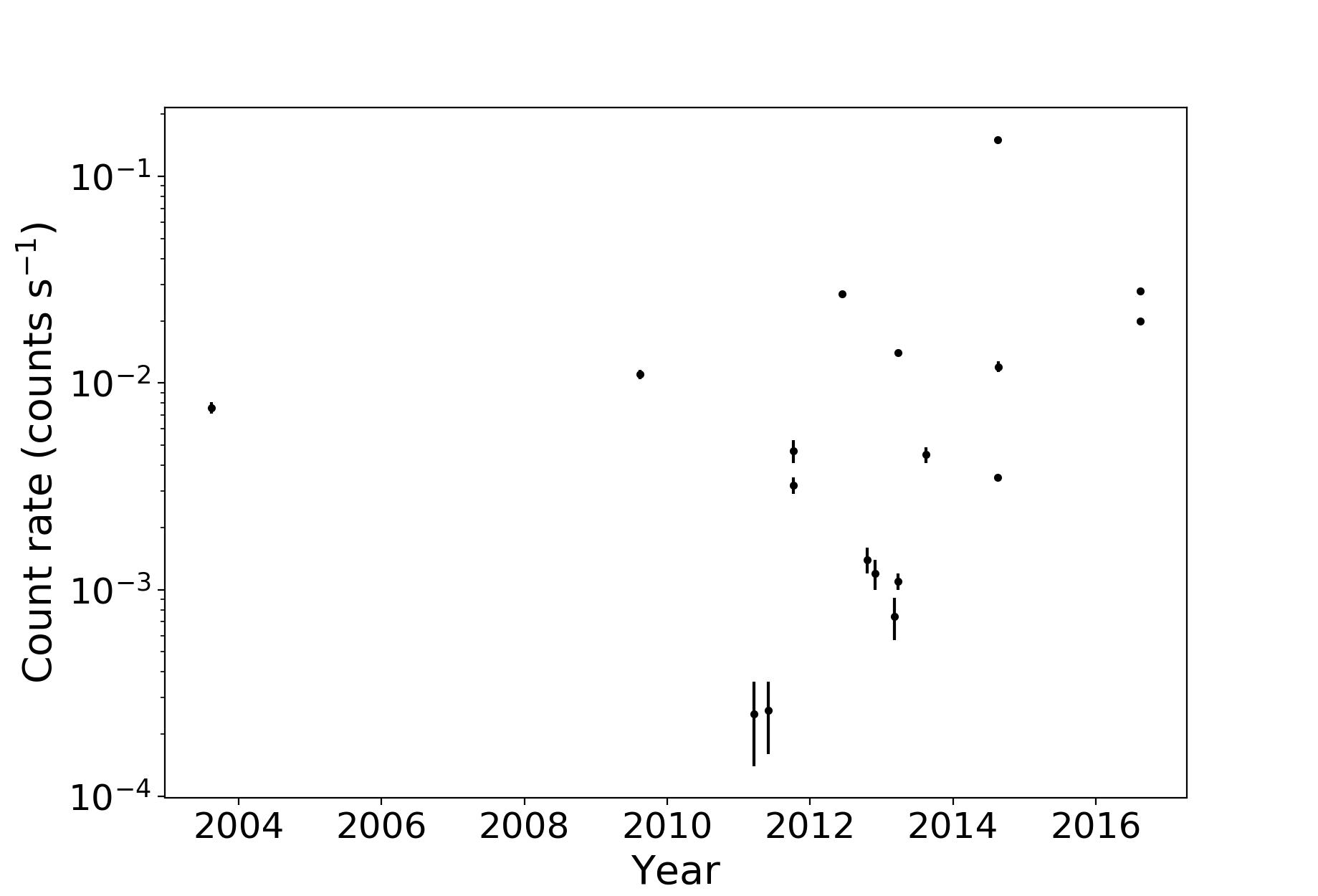}
 \caption{Light curve in the 0.5--10 keV energy range of our \textit{Chandra} observations of EXO
1745--248 during quiescence. The observations correspond to these described in Table \ref{chan_id}.}
\label{lc_obs}
\end{figure}

\begin{figure*}
\centering
\setlength{\fboxsep}{1pt}
  \begin{tabular}{@{}ccc@{}}
 \fbox{\includegraphics[width=.3\textwidth]{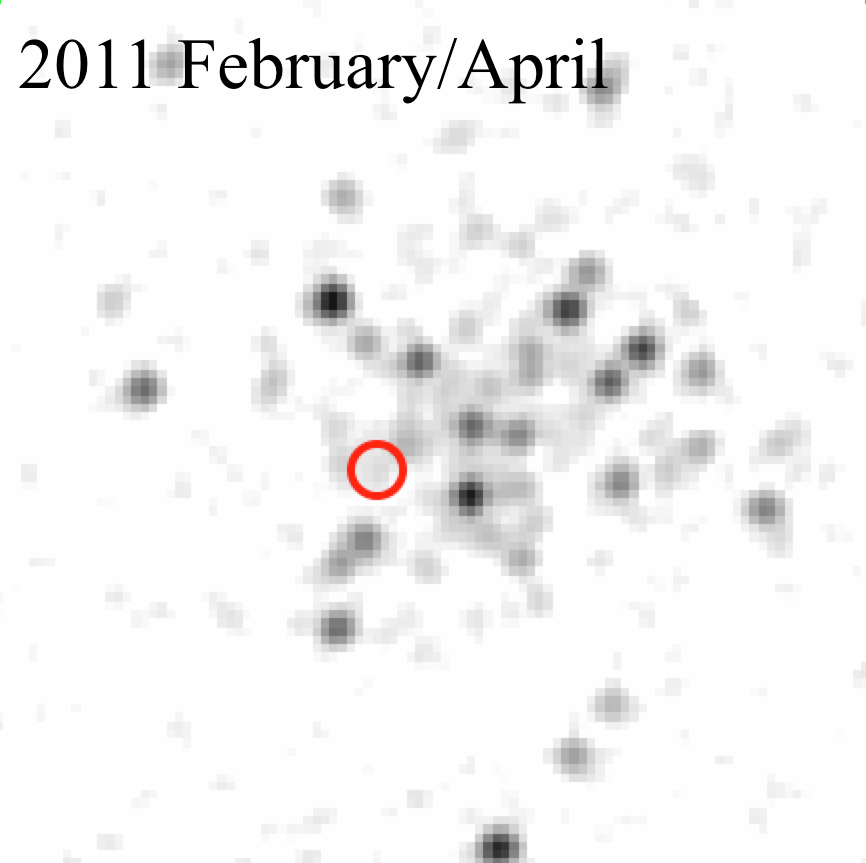}}&
 \fbox{\includegraphics[width=.3\textwidth]{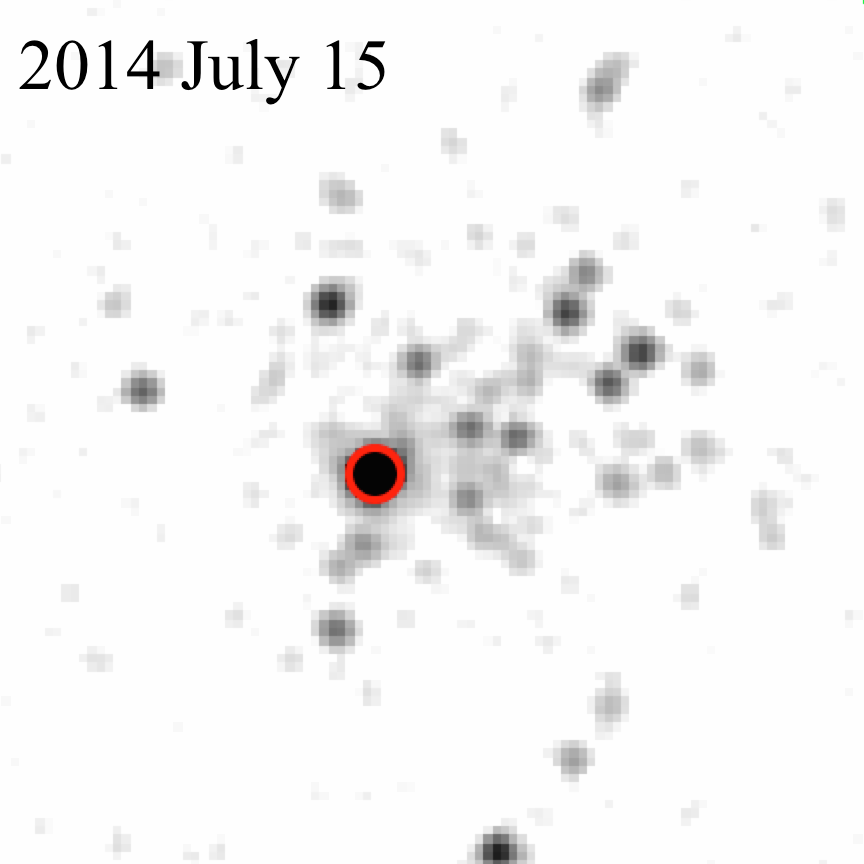}}&
 \fbox{\includegraphics[width=.3\textwidth]{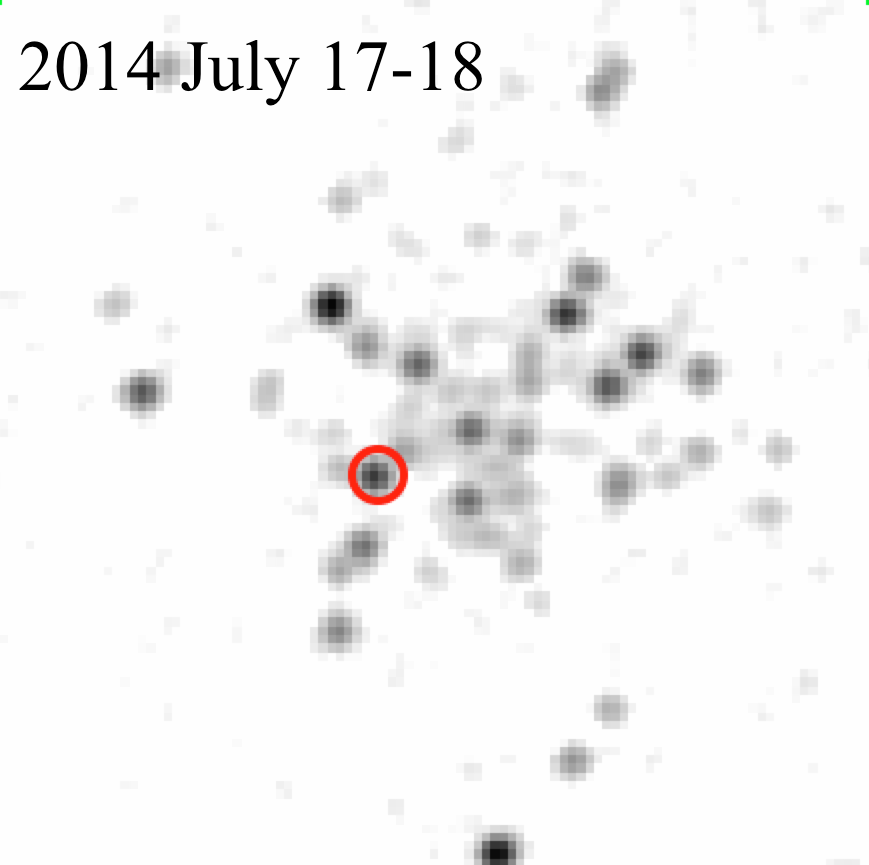}}\\
  \end{tabular}
  \caption{$Chandra$ ($50\arcsec \times 50\arcsec$) images  of Terzan 5 in the 0.5--7 keV energy range. The NS LMXB EXO 1745--248 is indicated with a red circle. Note the large variability of the source between different epochs. We show data obtained during observations in which the source was the faintest (left panel; Feb-Apr 2011; this is a stacked image of observations with IDs 13225 and 13252) and the brightest (middle panel; July 15, 2014). 
We also show the image obtained on July 17/18 (right panel), which was only taken two days after the brightest observation. Clearly the source was considerably fainter again, demonstrating that the source was highly variable on timescales of days (and not only on timescales of years).}
\label{fig:stacked}
\end{figure*}

\section{Observations and data reduction}

The $Chandra$ data used in this paper were taken at different epochs during the period 2003-2016 (see Table \ref{chan_id} for a log of the observations\footnote{Besides the $Chandra$ observations used in our paper, several additional $Chandra$ pointings have been performed on Terzan 5. However, during these observations one of the three transients located in this GC was active and their bright luminosities strongly influenced the data quality for faint sources. Therefore, we do not discuss these observations in this paper.}). The data were downloaded from the $Chandra$ archive\footnote{http://cxc.harvard.edu/cda/}. 
All the data were taken in the faint mode, with a nominal frame time of 3.2 s. 
The target was positioned on the S3 chip. 

For the data reduction we have followed the $Chandra$ threads\footnote{http://cxc.harvard.edu/ciao/threads/} which make use of the \caps{ciao} software package \citep[v4.9;][]{2006fru}. 
We recalibrated all the data using the $chandra\_repro$ script to assure that the newest calibrations were applied. 

To search for potential periods of high background (likely due to background flares), for each observation we created background light curves excluding the source region. No episodes of high background rates were observed for all but one observation: the data set with identification number (ObsID) 3798. 
Therefore, we reprocessed that data set following the appropriate $Chandra$ tools\footnote{http://cxc.harvard.edu/ciao/threads/filter/}
to remove episodes of high background (i.e., the last 8.1 ks of the observation were removed). This reduced its exposure time to 31.2 ks. 
The exposure times for all used observations are given in Table \ref{chan_id}.

To create the left panel of Figure \ref{fig:stacked}, we stacked ObsID 13225 and 13252 during which the source was particular faint. To correctly combine these data sets, we first carried out relative astrometry (besides the absolute astrometry provided by $Chandra$) following the appropriate $Chandra$ thread\footnote{http://cxc.harvard.edu/ciao/threads/reproject$\_$aspect/}, and we created a broad band (0.5--7 keV) source catalog with the routine \caps{fluximage}. The corresponding PSF map was created with \caps{mkpsfmap}, while the \caps{wavdetect} algorithm was used to locate the sources. The task $wcs\_match$ was used to perform a source cross match. This routine also determines the transformation parameters to shift a given image to the reference image (ObsID 15615).

\begin{table}
\begin{center}
\caption{Log of the $Chandra$ observations of EXO 1745--248 used in this paper. The net, background-corrected count rate given in column 4 corresponds to the energy band  0.5--10 keV.
The count rates are averages for the whole observation.}
\begin{tabular}{ l | c | c |  l }
\hline
Date &  Obs ID & Exposure & Net count rate \\
        &              &  time (ks)  & (counts s$^{-1}$)\\
\hline
2003-07-13/14       & 3798 &  31.2$^1$ & $7.6 \pm 0.5 \times 10^{-3}$\\
2009-07-15/16        & 10059 &   36.3  &$1.12\pm0.06 \times 10^{-2}$ \\ 
2011-02-17	&13225& 29.7 & $2.5\pm1.1\times 10^{-4}$\\
2011-04-29/30	&13252& 39.5 & $2.6\pm1.0\times 10^{-4}$\\
2011-09-05      &13705& 13.9 & $4.7\pm 0.6\times 10^{-3}$\\
2011-09-08	&14339& 34.1 & $3.2\pm 0.3\times 10^{-3}$\\
2012-05-13/14	&13706& 46.5 & $2.79\pm 0.08\times 10^{-2}$\\
2012-09-17/18	&14475& 30.5 & $1.4\pm 0.2\times 10^{-3}$\\
2012-10-28	&14476&  28.6 & $1.2\pm 0.2\times 10^{-3}$\\
2013-02-05        &14477& 28.6 & $7.4\pm 1.7\times 10^{-4}$\\
2013-02-22	&14625& 49.2 &$1.44\pm 0.05\times 10^{-2}$\\
2013-02-23/24	&15615& 84.2 & $1.1\pm 0.1\times 10^{-3}$\\
2013-07-16/17	&14478&  28.6 & $4.5\pm 0.4\times 10^{-3}$\\ 
2014-07-15        &14479& 28.6 & $1.48\pm0.02\times 10^{-1}$\\
2014-07-17/18	&16638& 71.6 & $3.49\pm 0.02\times 10^{-3}$\\
2014-07-20	&15750& 23 & $1.16\pm0.07\times 10^{-2}$ \\
2016-07-13/14	&17779& 68.9 & $2.84\pm0.06\times 10^{-2}$\\
2016-07-15/16	&18881&64.7 &  $2.04\pm0.06\times 10^{-2}$\\
\hline
\end{tabular}
{
\flushleft
\startfoot  
\textfoot{Effective exposure time after the subtraction of background flares. The original exposure time was 39.3 ks. }
}
\label{chan_id}
\end{center}
\end{table}

The position of EXO 1745--248 has been previously constrained using $Chandra$ by identifying the source in outburst \citep{2003heinke}, resulting in a sub-arcsecond precision position of our target. 
This allows us to straightforwardly identify the right quiescent counterpart among the other faint cluster sources (see Figure \ref{fig:stacked}).
To extract light curves and spectra of EXO 1745--248, we used circular regions with variable radius centered on the source.
We used an extraction region with a radius ranging from $\sim1\arcsec$ to $\sim1.7\arcsec$ depending on the (variable) source flux (see Figure \ref{fig:stacked} and Section \ref{analysis}).
The background spectra were extracted from a source free part of the CCD using a circular region with a radius of $10\arcsec$.
We used the task \caps{dmextract} for creating light curves in the energy range 0.5--7 keV.
The time bins were optimized (see Table \ref{free}) to avoid as much as possible bins with zero counts.

The tool \caps{specextract} was
used to obtain the source and background spectra (using the same extraction regions as used for the light curve extraction) and to generate the ancillary response
files and redistribution matrix files. The data were grouped to have at least 15 counts per bin for most of the observations. 
This allows the use of the $\chi^2$ minimization fitting technique.
However, for data sets in which the source was very faint, we rebinned the data 
to have at least 1 count per bin. For these observations we used the background subtracted Cash statistics \citep[W-statistics;][]{1979wa} instead of $\chi^2$ (see Table \ref{free}). The spectra were grouped using the tool \caps{dmgroup}.

To carry out all the spectral fitting we used the software package \caps{Xspec} \citep[v12.9.1][]{1996ar}.
To model the hydrogen column density N$_H$ we used the model $tbnew\_gas$\footnote{http://pulsar.sternwarte.uni-erlangen.de/wilms/research/tbabs/}  with WILM abundances \citep{2000wilm} and VERN cross-sections \citep{1996ver}.
We fitted our spectra with a power-law model (\textsc{pegpwrlw}). In one case (ObsID 14479; Section \ref{thermal}), we also used a two component model consisting of a power-law model plus a black-body model (\textsc{bbodyrad}).
The luminosities were calculated using a distance towards Terzan 5 of $5.5\pm0.9$ kpc \citep{2007orto}.
All errors quoted in the paper correspond to 90\% confidence levels.

\section{Results}

\subsection{Spectral analysis and variability}
\label{analysis}

To study the quiescent spectral behavior of EXO 1745--248 we analyzed its X-ray spectrum at different epochs in the energy band 0.5-10 keV. 
\citet{2005wijnands} and \citet{2012dege} carried out similar spectral studies of EXO 1745--248 in 
quiescence using a sub-set of the observations we are presenting here.
They found that the spectra can be well described by a power-law model, and that the inclusion of a thermal component did not improve the spectral fits.

\begin{figure}
\centering
\vspace{-20pt}
  \includegraphics[width=.5\textwidth, trim=1.5cm 0.0cm 0.0cm 0cm]{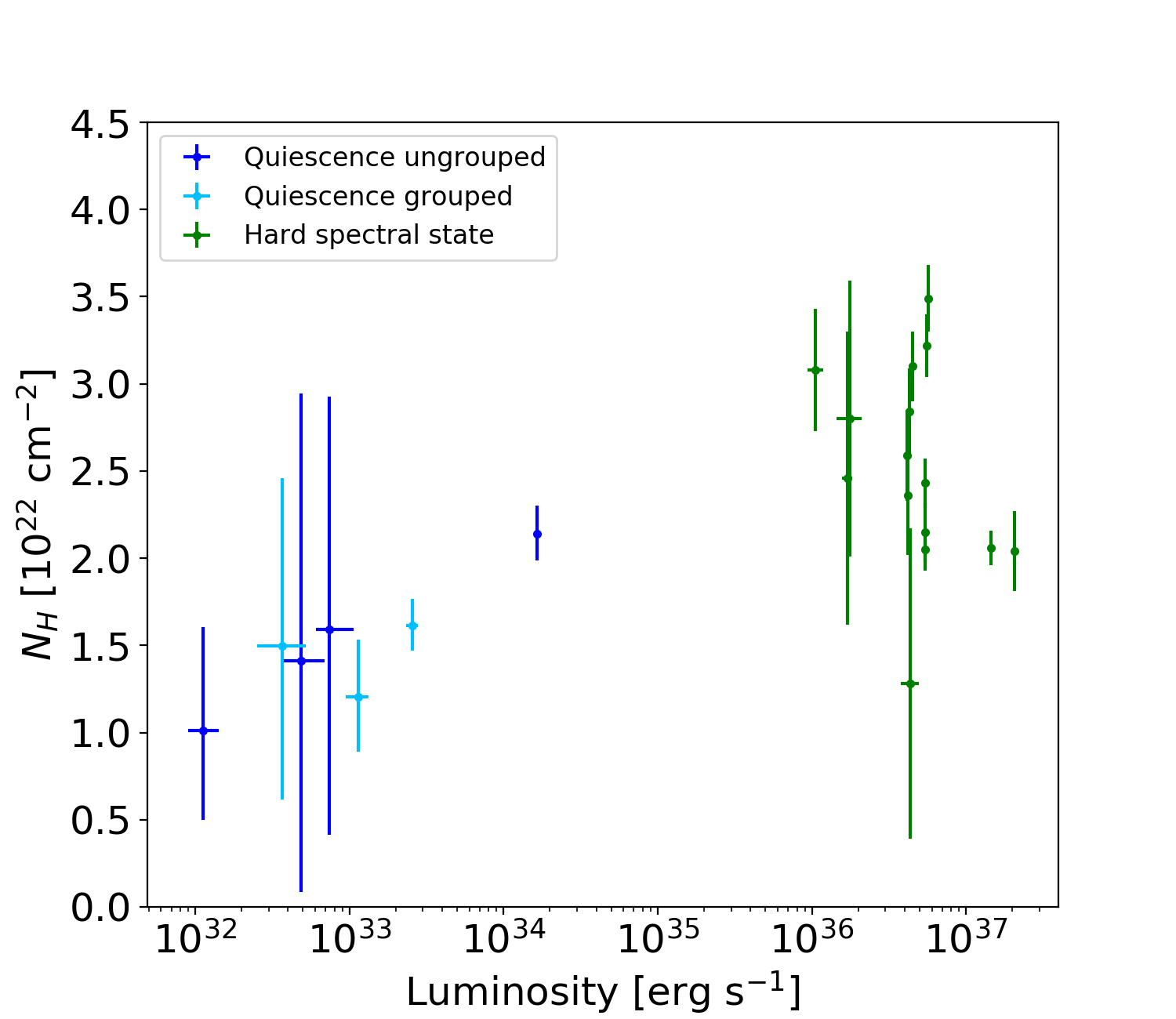}
 \caption{The hydrogen column density ($N_H$) versus luminosity (0.5--10 keV) for EXO 1745--248  for the case in which $N_H$, $\Gamma$ and $F_X$
are free to vary for each individual spectrum. The blue (dark and light) points correspond to the quiescent state as analyzed in this work. Data (the light blue points) are grouped as in Table \ref{free}. The two ungrouped dark blue points below $5\times10^{32}$ erg/s were obtained using W-statistics. The rest of the quiescent points were obtained using $\chi^2$. 
The green points correspond to the very hard state data points of the source published by \citet{2017parikh}.}
\label{all_free}
\end{figure}

Based on these studies, we also used an absorbed power-law model (\textsc{pegpwrlw} in \caps{Xspec}).
In this model the hydrogen column density ($N_H$), the photon index ($\Gamma$) and the X-ray flux ($F_X$\footnote{Which corresponds to the power-law normalization in the pegpwrlw model.}) are the model parameters. 
Initially we left all of them free to vary in the fits (for all the spectra; see Table \ref{free} for the results of these fits).
However, for several data sets that have a small number of net counts, we noted that $N_H$ and $\Gamma$ have large errors or that the fit did not converge. This suggests that 
the signal-to-noise ratio (S/N) of our spectra is too low to constrain the parameters of this simple, phenomenological, model. In these cases we fixed the values of $N_H$ and $\Gamma$ to $1.4\times 10^{22}$  cm$^{-2}$ and 1.4, respectively. 
These values correspond to the average $N_H$ and $\Gamma$ derived from the observations 
for which the quality of the spectra was good enough to obtain reasonably constrained parameter values (these are the observations for which the $\chi^2$ statistics could be used). The obtained results are also shown in Table \ref{free}.

\begin{figure}
\centering
\includegraphics[width=.5\textwidth, trim=1.cm 0.0cm 0.0cm 0cm]{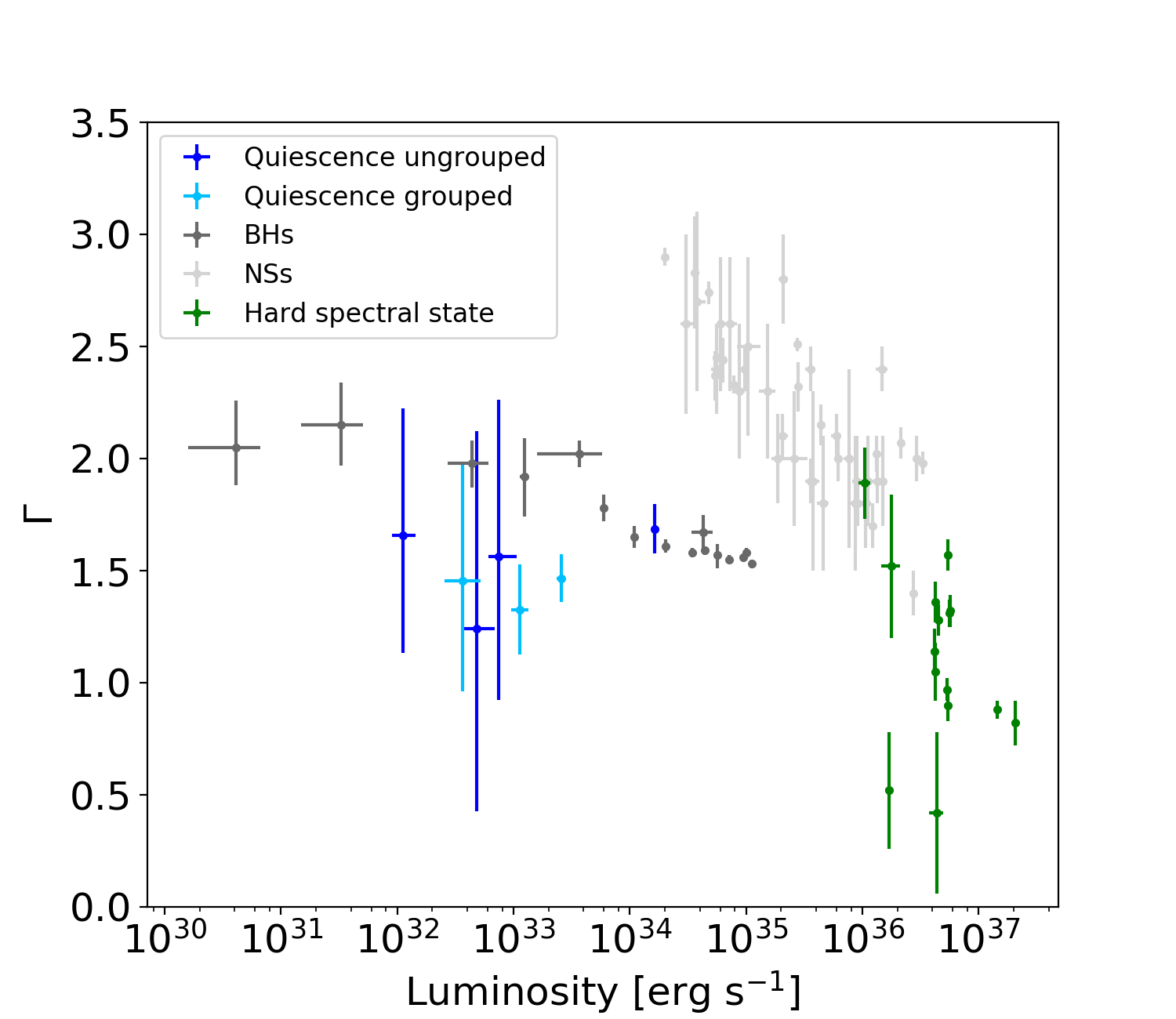}
\caption{The photon index $\Gamma$ versus luminosity (0.5--10 keV) for EXO 1745--248 in the case in which $N_H$ is allowed to vary for each individual spectra. The blue (dark and light) points are our quiescent points. The green points correspond to the very hard state \citet{2017parikh}. 
For comparison, the results for other NS systems (light gray points) and for BH systems (dark gray points) are also plotted \citep[taken from][]{2015wij}.
The grouped light blue points correspond to the data groups given in Table \ref{free}. The two ungrouped dark blue points below $5\times10^{32}$ erg/s were obtained using W-statistics. The rest of the quiescent points were obtained using $\chi^2$. 
}
\label{gama_lx_free}
\end{figure}

For the observations performed in 2003 and 2009 we obtained results consistent with those reported by \citet{2005wijnands} and \citet{2012dege}. For ObsID 13225 and 13252 (February and April 2011), \citet{2012dege} imposed $N_H=1.2 \times10^{22}$ cm$^{-2}$ and $\Gamma=1.5$ and found a 0.5--10 keV X-ray luminosity $L_X \sim 4\times10^{31}$ erg s$^{-1}$. This is consistent with our results, even though we have used different abundances and a slightly different $N_H$ absorption model.

We note that at a given X-ray luminosity, the spectral parameters are constant within uncertainties (e.g. compare ObsID 10059 and 15750).
To better constrain the spectral parameters, we therefore combined several of these observations into groups according to their luminosity. 
For each group we tied $N_H$ and $\Gamma$, but we left $F_X$ free to vary since it was found to slightly change when fitting the observations individually. We then fitted each group separately. The results of these fits are also displayed in Table \ref{free}.

In Figure \ref{all_free}, we compare $N_H$ from this work with that of \citet{2017parikh} of EXO 1745--248 during its 2015 outburst ($L_X>1\times10^{36}$ erg s$^{-1}$). From this figure it can be seen that the $N_H$ during outburst reached higher values (although not always) than what we observe in quiescence. In addition, there is the indication of a trend during quiescence, with $N_H$ also increasing with the flux. This may suggest that the absorption internal to the system increases. Note that the obtained values of $N_H$ are close to these of the cluster itself  \citep[][]{2014bara,2015bara}, indicating that the contribution of any additional $N_H$ absorption component (if any) would be minimal.

In Figure \ref{gama_lx_free}  we show $\Gamma$ versus $L_X$ and compare our results with that of EXO 1745--248 during its 2015 outburst and with that of a large group of NS and BH systems \citep[Figure 1 of][]{2015wij} during outburst (BH and NS) and those that meet our definition of quiescence (BH only). We clearly see that the photon index of EXO 1745--248 is $\sim1.5$ over the full 2 orders of magnitude range of the quiescent luminosity (10$^{32}-10^{34}$ erg s$^{-1}$). Moreover, the photon index is lower, and thus the source is harder than BH systems at similar luminosities. This remains true even if we scale the results to Eddington luminosities. Interestingly, $\Gamma$ is even lower (0.5--2) above $1\times10^{36}$ erg s$^{-1}$. This led \citet{2017parikh} to suggest that during its 2015 outburst the source was in a newly identified very hard state. Currently, we have no high S/N data for the luminosity range $\sim1\times10^{34}$ to $\sim1\times10^{36}$ erg s$^{-1}$, so it is unclear if the source follows the general NS trend and thus becomes softer below $1\times10^{36}$ erg s$^{-1}$, but then suddenly hardens around $1\times10^{34}$ erg s$^{-1}$, or if it stayed very hard over the full luminosity range covered in this figure.

\begin{figure}
\centering
\vspace{33pt}
  \includegraphics[width=.5\textwidth, trim=0.5cm 0.5cm 1.0cm 2.2cm]{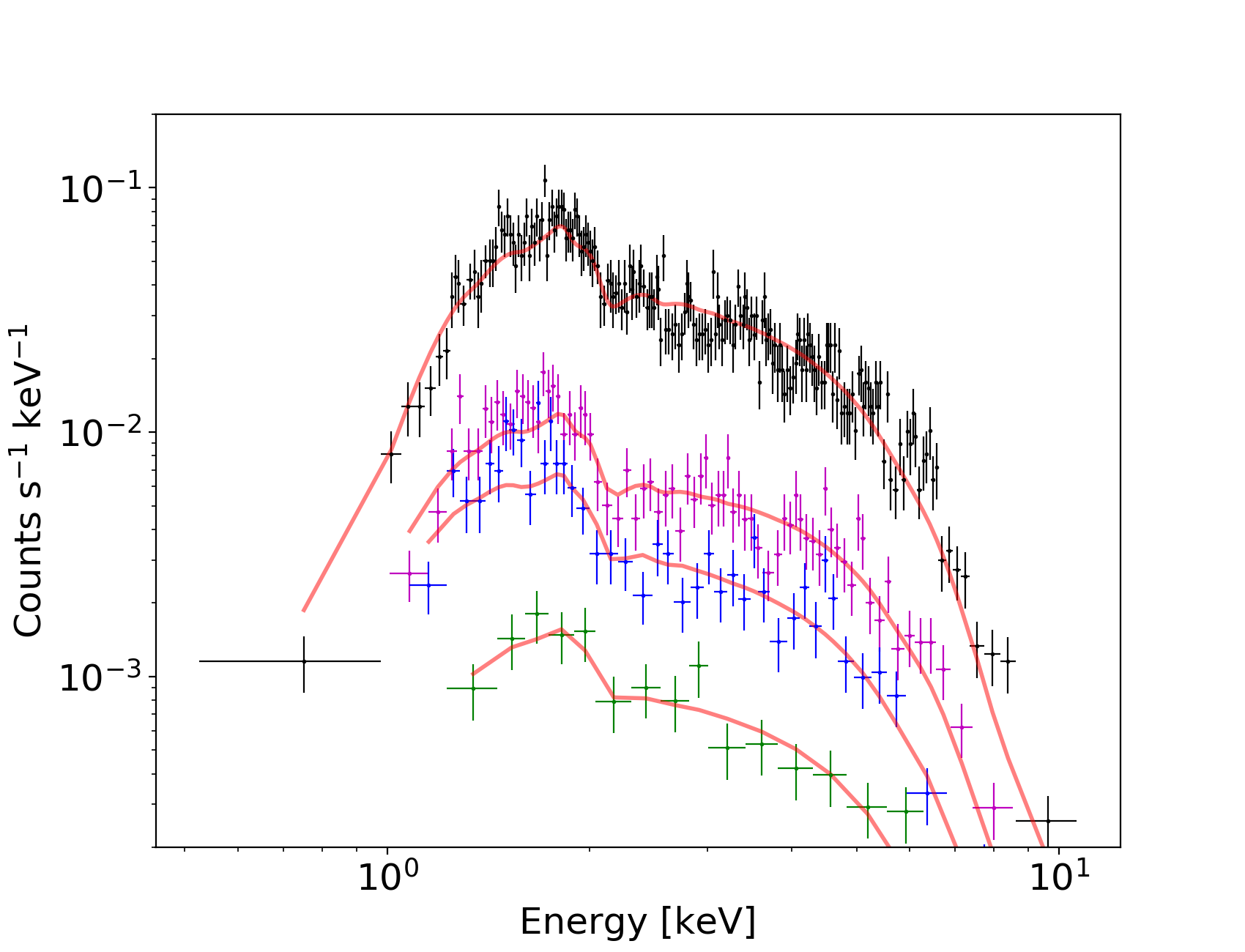}
 \caption{The $Chandra$ spectra of EXO 1745--248 observed at different epochs during its quiescent state. 
The energy range is 0.5-10 keV.
From top to bottom, the spectra obtained during the observations with the following ObsIDs are plotted: 14479, 13706, 14625 and 16638. 
X-ray luminosities in the range $4\times10^{32}-1.7\times10^{34}$ erg s$^{-1}$ are observed between these observations.
In each of these observations the system shows luminosity variations of around two orders of magnitude in timescales of a few hours (see Figure \ref{fig:lightcurves1}).
The solid lines represent the best absorbed power-law model in which $N_H$, $\Gamma$ and $F_X$ were free parameters in the fit.}
\label{multiple_spec}
\end{figure}

In Figure \ref{multiple_spec} we show the spectra of ObsIDs 14479, 13706, 14625 and 16638, taken during the period 2012--2014, 
where the strong variability of the object is clearly visible. The X-ray luminosity of EXO 1745--248 was found to vary from $4\times 10^{32}$ erg s$^{-1}$  up to $\sim1.7\times 10^{34}$ erg s$^{-1}$ (see Table \ref{free}). This shows that during quiescence, the object exhibits luminosity variations of several orders of magnitude on timescales of days to years (see also Figure \ref{fig:stacked} and Table \ref{chan_id}).

To investigate the variability of the source in more detail, we created light curves in the energy range 0.5-10 keV (Figure \ref{fig:lightcurves1}).  
Due to the large difference in source count rates during the individual observations, we used a variety of time bin sizes (see Table \ref{free}) to highlight the variability, trying to avoid as much as possible time bins with count rates of zero. We do not plot the light curve of ObsID 13252 because of the very low number of counts.

\begin{figure*}
\centering
  \begin{tabular}{@{}cc@{}}
 \includegraphics[width=.45\textwidth,   trim =1.5cm 0.0cm 1.0cm 0cm]{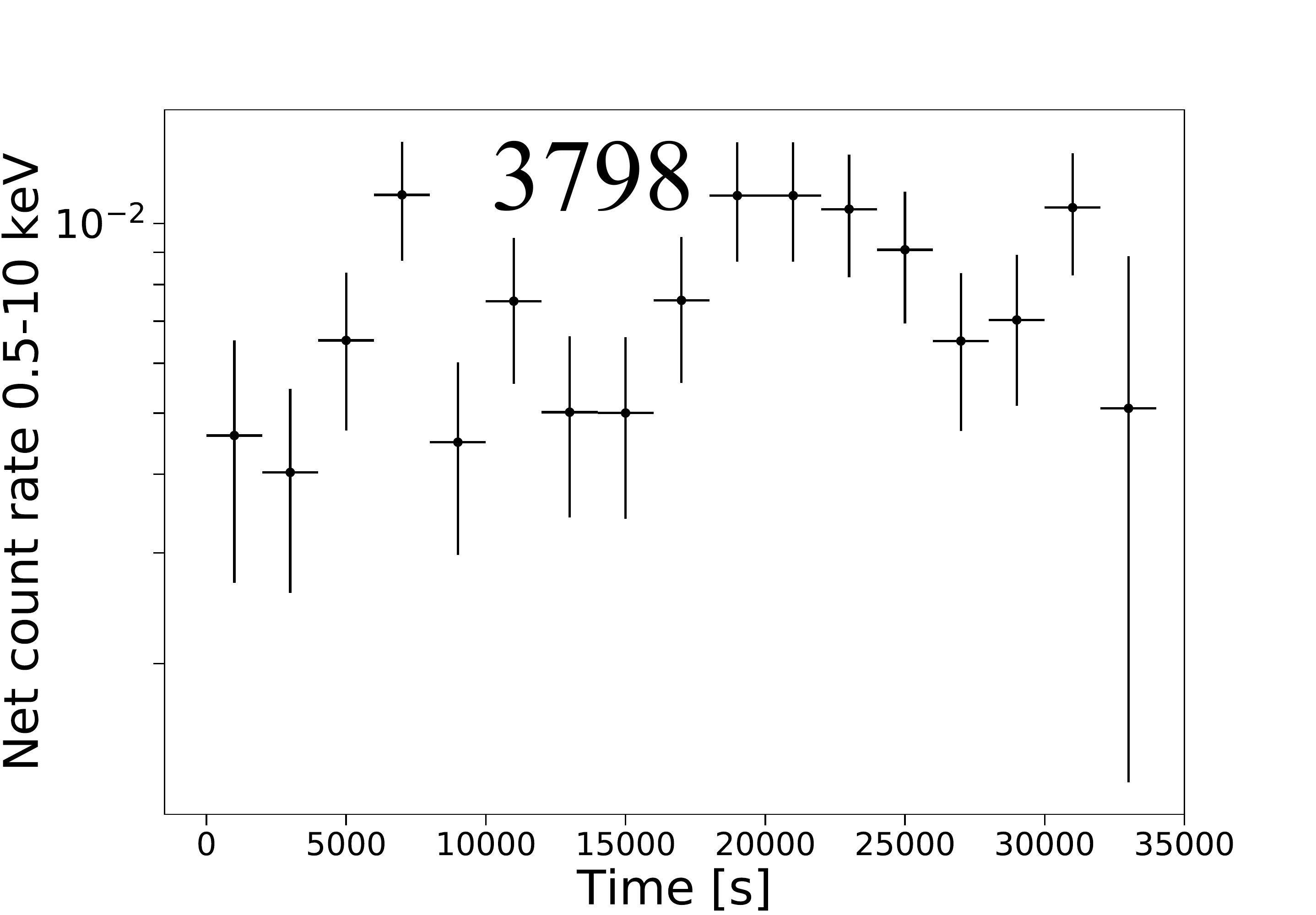}&
 \includegraphics[width=.45\textwidth,   trim =1.5cm 0.0cm 1.0cm 0cm]{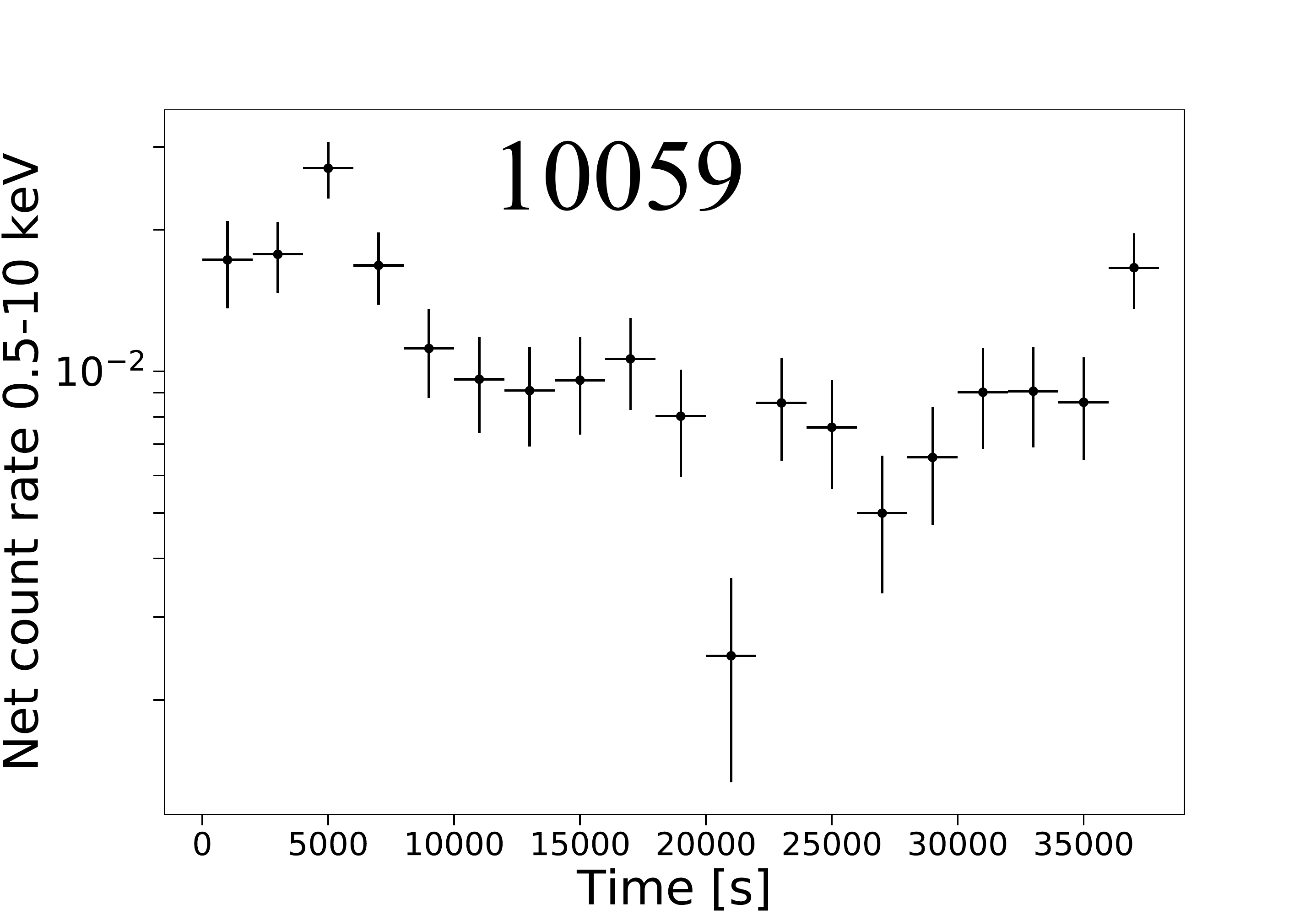}  \\
 \includegraphics[width=.45\textwidth,   trim =1.5cm 0.0cm 1.0cm 0cm]{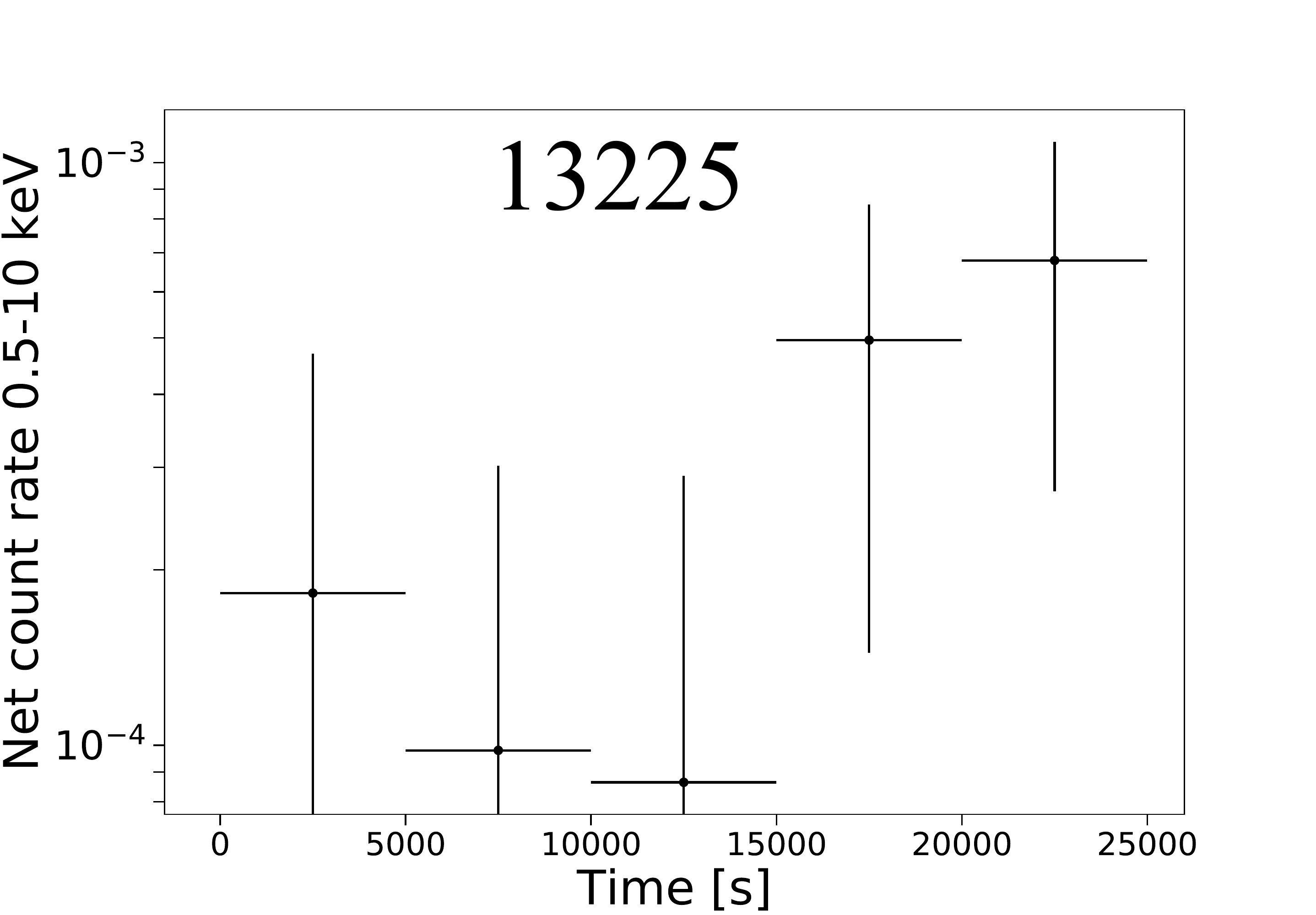}&
 \includegraphics[width=.45\textwidth,   trim =1.5cm 0.0cm 1.0cm 0cm]{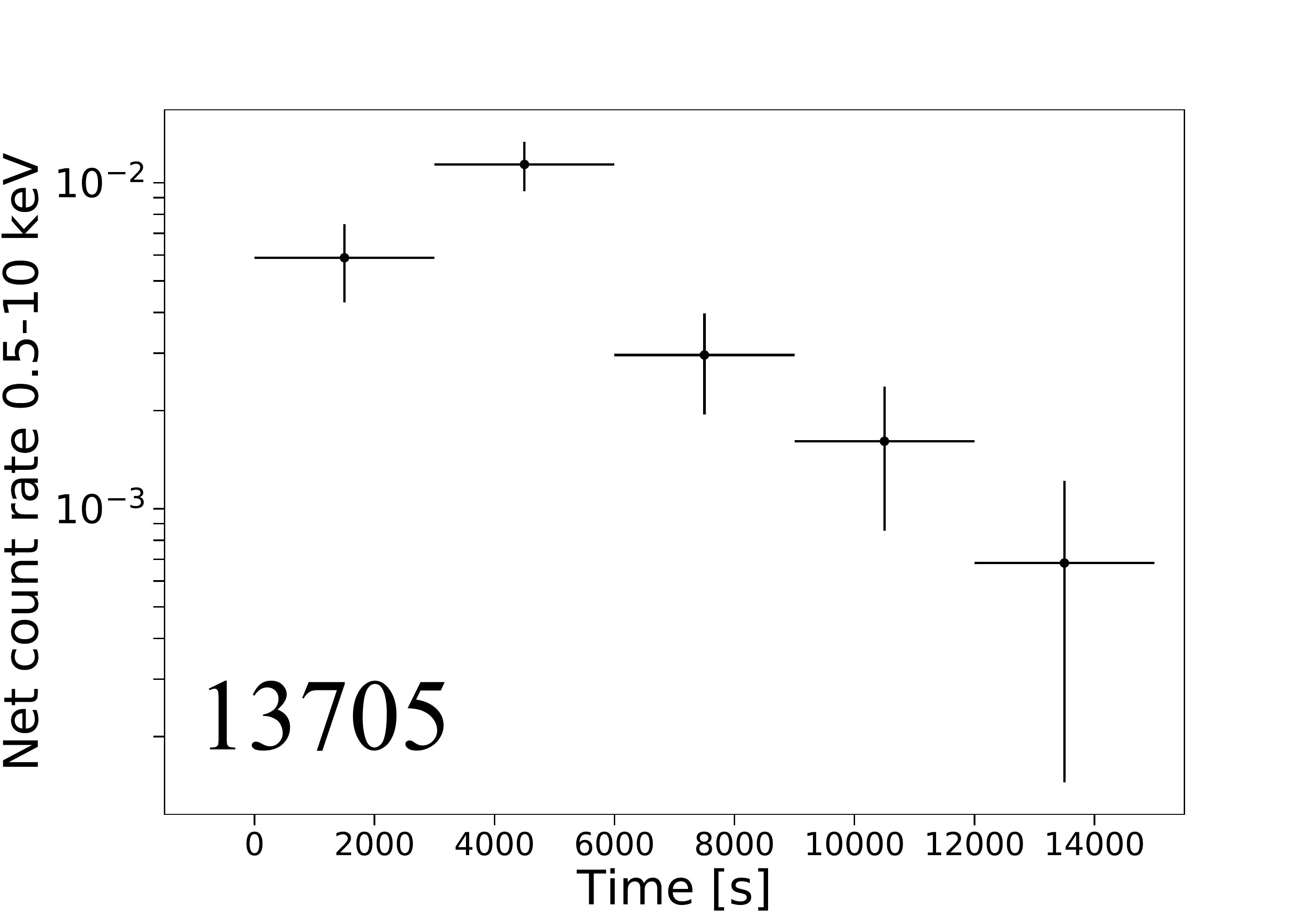}  \\
 \includegraphics[width=.45\textwidth,   trim =1.5cm 0.0cm 1.0cm 0cm]{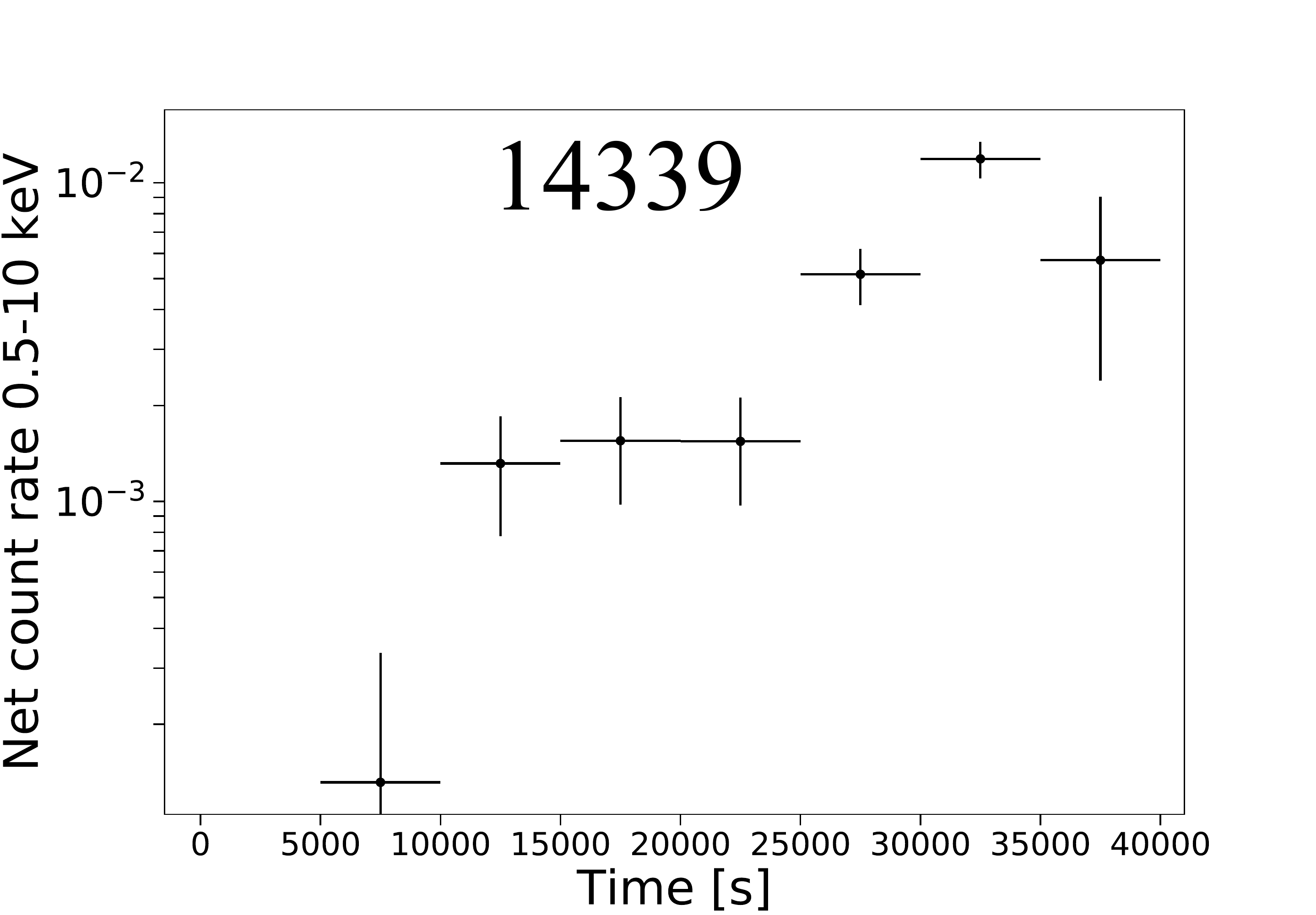}&
 \includegraphics[width=.45\textwidth,   trim =1.5cm 0.0cm 1.0cm 0cm]{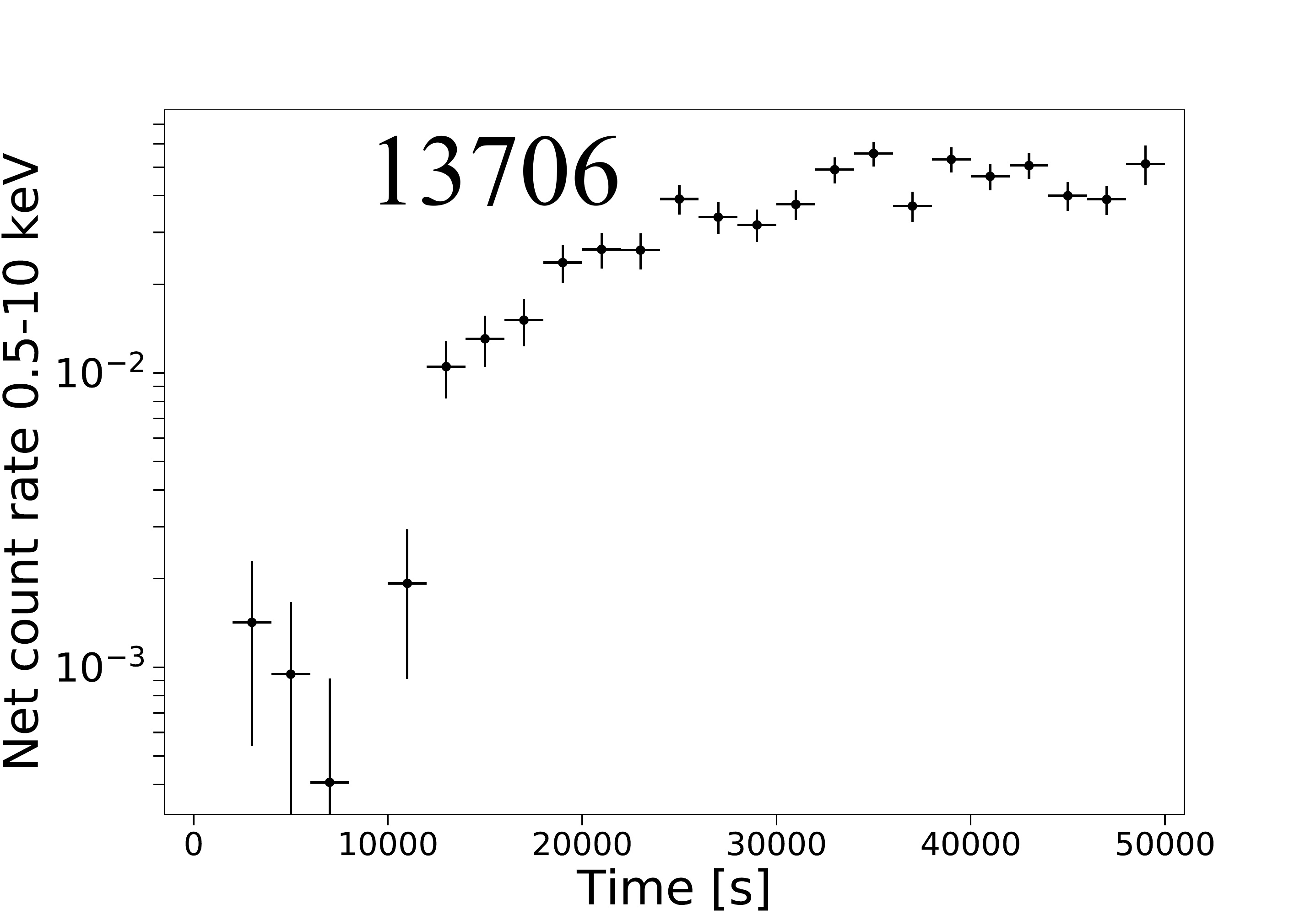}  \\
  \end{tabular}
  \caption{Light curves of EXO 1745--248 in the 0.5-10 keV X-ray band during quiescence. The ObsIDs correspond to the different $Chandra$ observations listed in Table \ref{chan_id}. They are ordered in chronological order from left to right, top to bottom.
Only the lightcurve for ObsID 13252 is not shown due to the small number of counts.
For each ObsID we used a different bin size to highlight the variability but ensuring that we have as much as possible no bins with zero photons.
Reference time is the start time of each observation.}
\label{fig:lightcurves1}
\end{figure*}

\begin{figure*}
\centering
  \begin{tabular}{@{}cc@{}}
 \includegraphics[width=.45\textwidth,   trim =1.5cm 0.0cm 1.0cm 0cm]{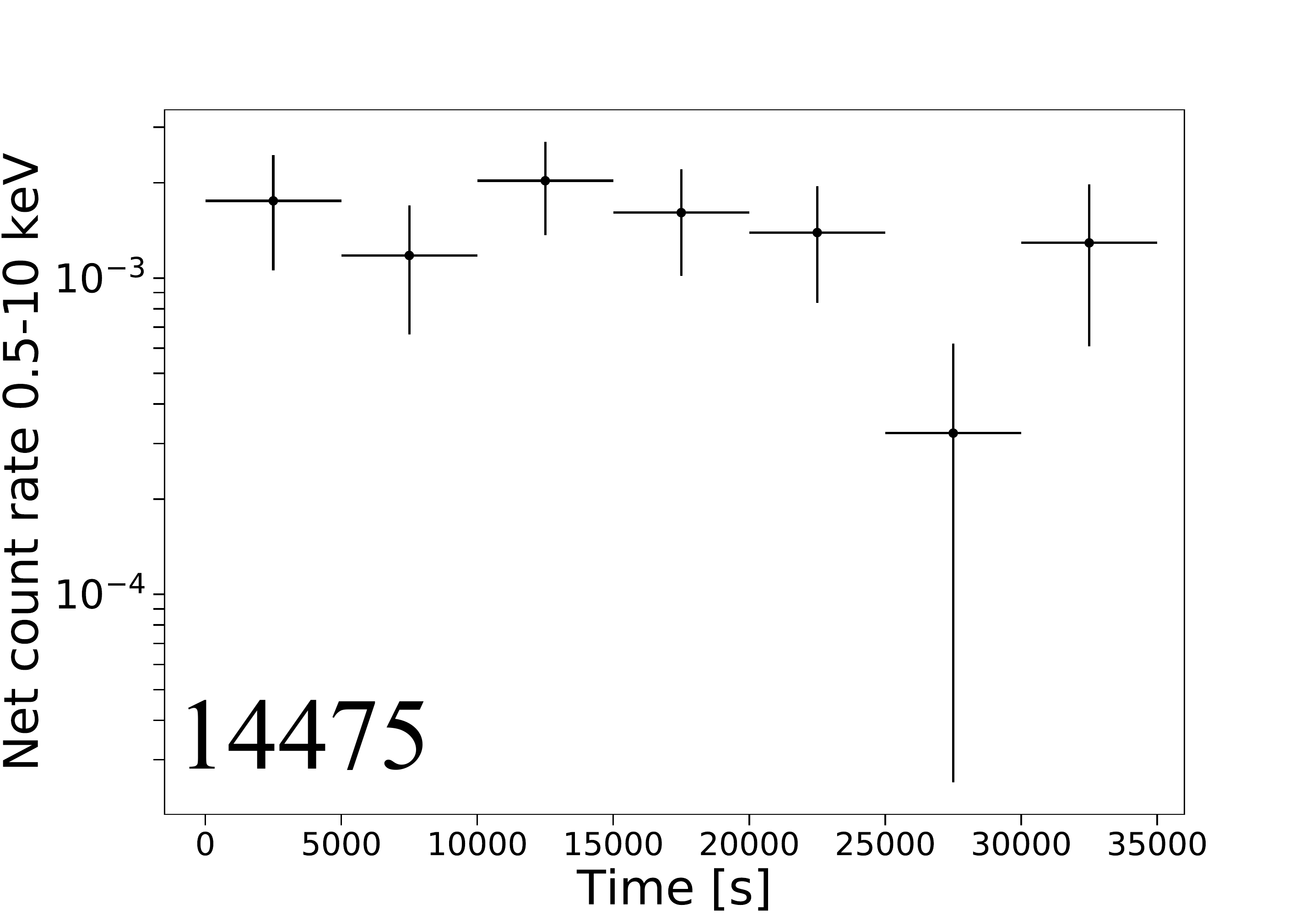}&
 \includegraphics[width=.45\textwidth,   trim =1.5cm 0.0cm 1.0cm 0cm]{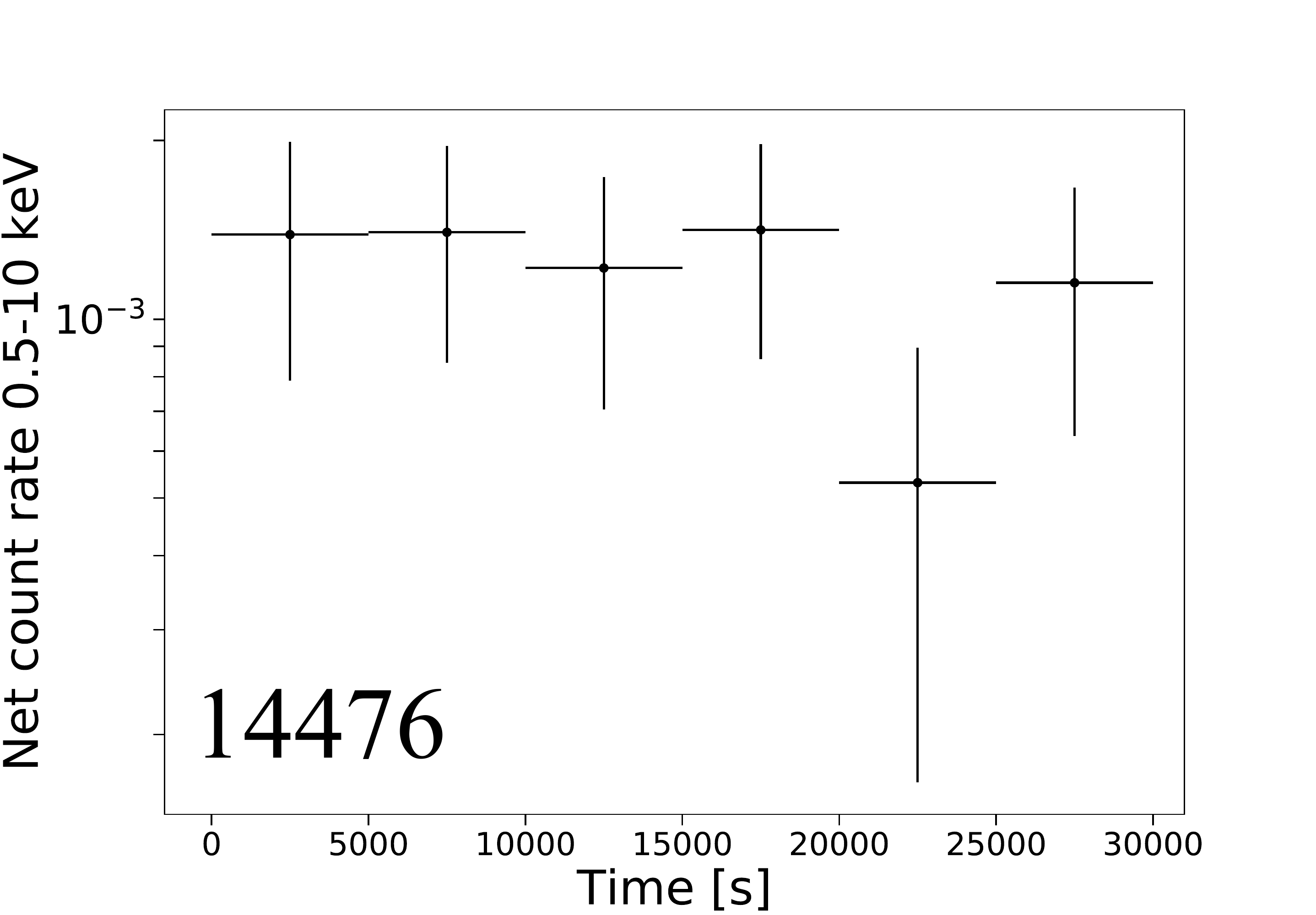}  \\
 \includegraphics[width=.45\textwidth,   trim =1.5cm 0.0cm 1.0cm 0cm]{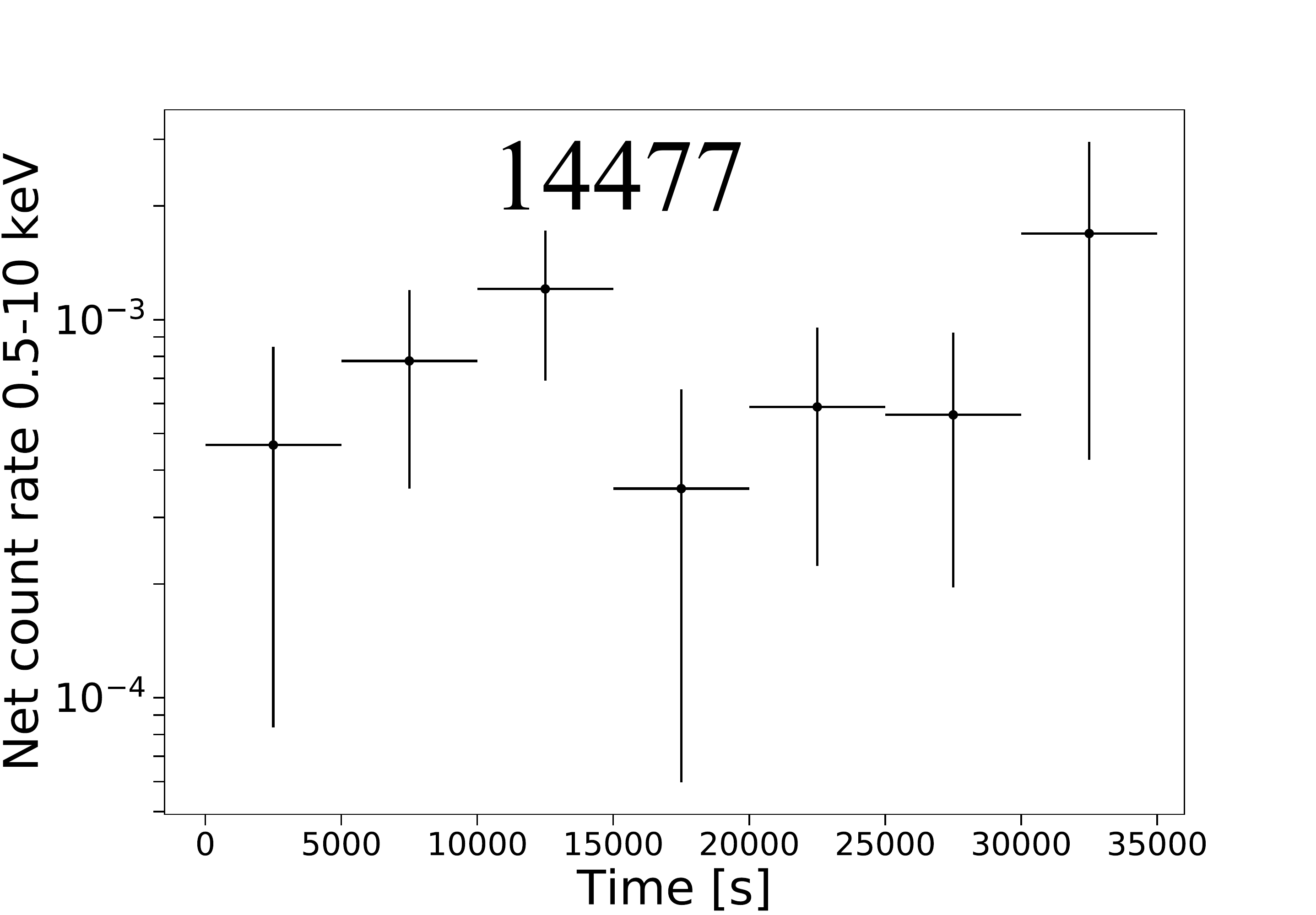}&
 \includegraphics[width=.45\textwidth,   trim =1.5cm 0.0cm 1.0cm 0cm]{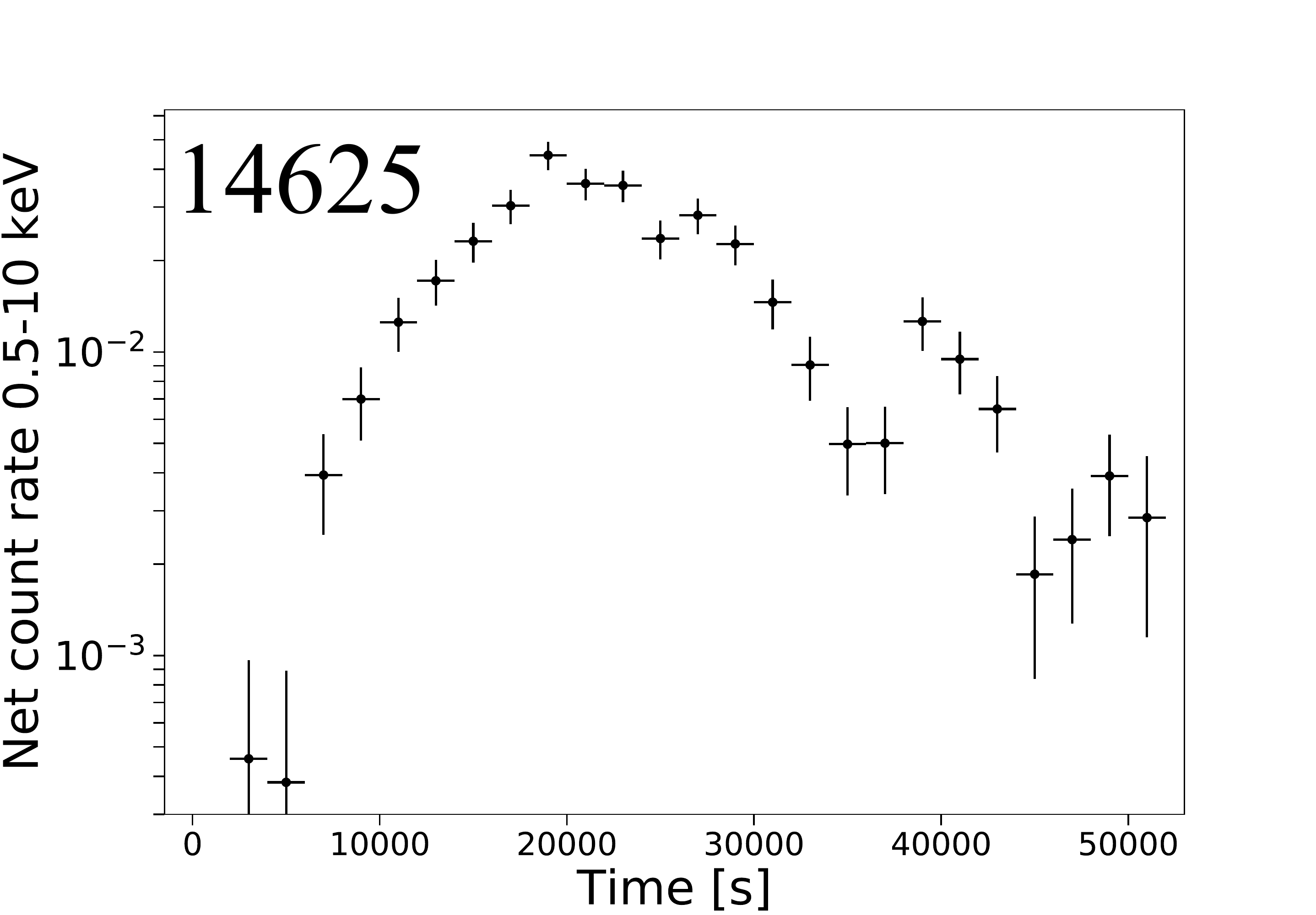}  \\
 \includegraphics[width=.45\textwidth,   trim =1.5cm 0.0cm 1.0cm 0cm]{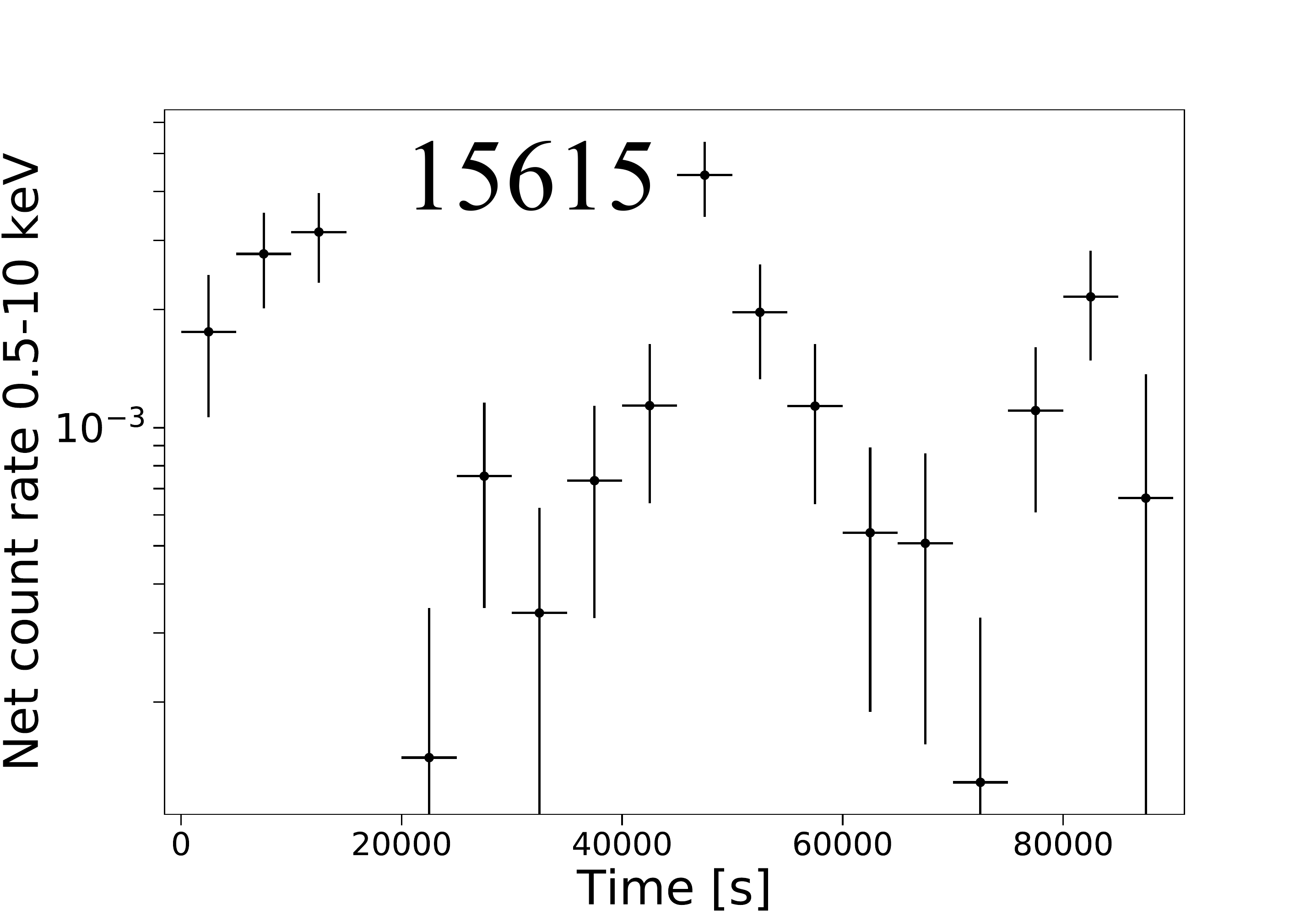}&
 \includegraphics[width=.45\textwidth,   trim =1.5cm 0.0cm 1.0cm 0cm]{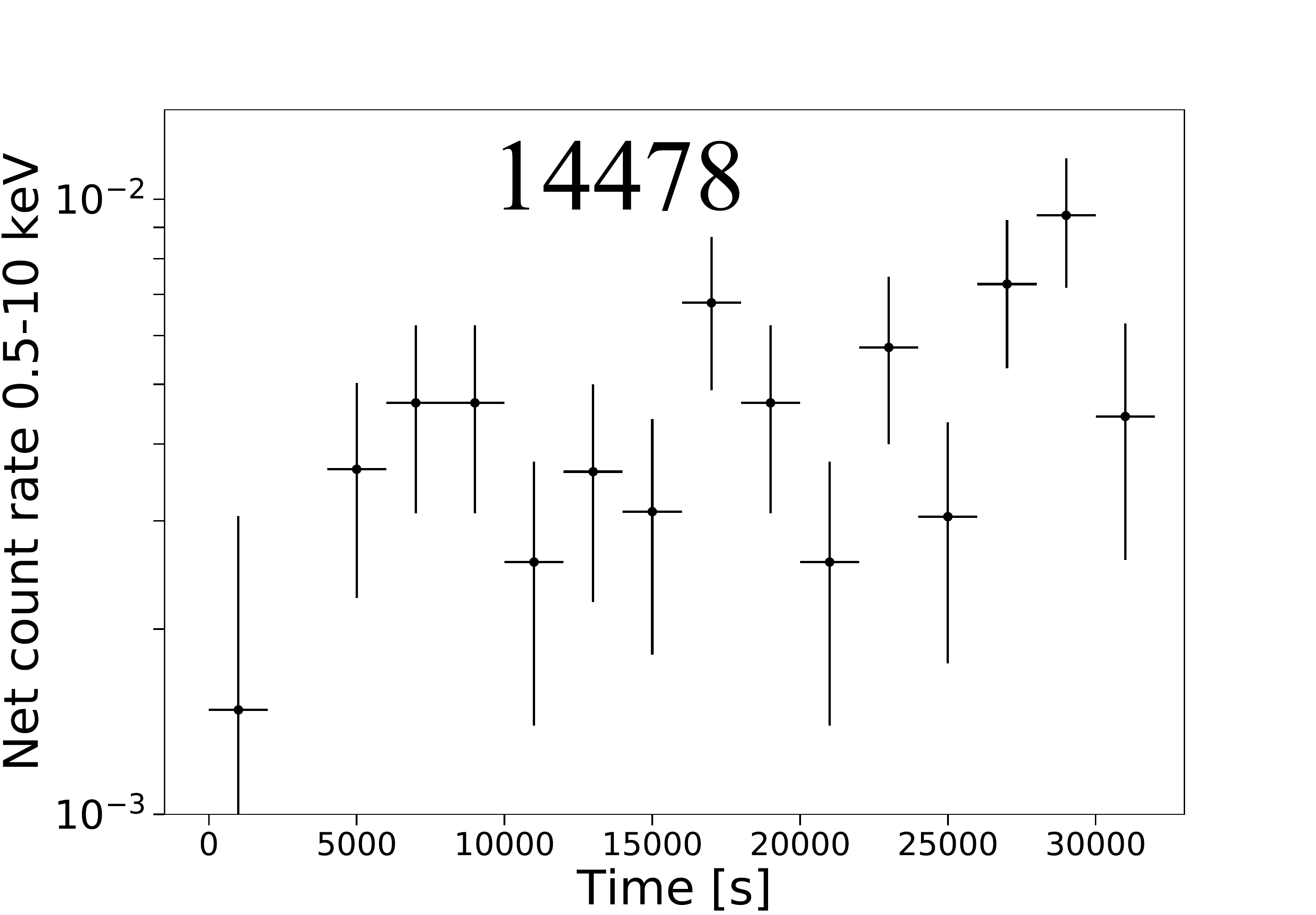}  \\
  \end{tabular}
\addtocounter{figure}{-1}
  \caption{Continued}
\end{figure*}

\begin{figure*}
\centering
  \begin{tabular}{@{}cc@{}}
 \includegraphics[width=.45\textwidth,   trim =1.5cm 0.0cm 1.0cm 0cm]{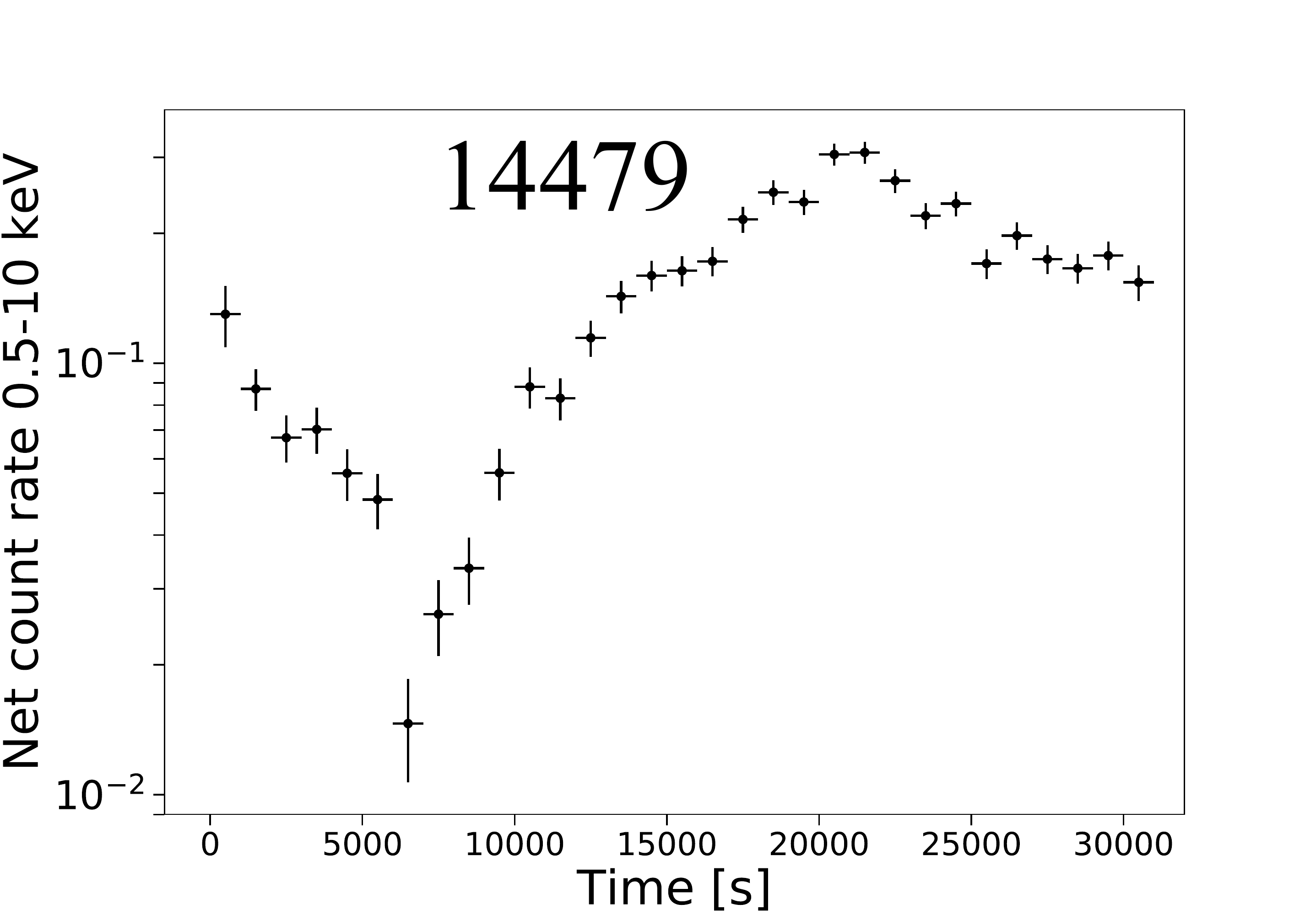}&
 \includegraphics[width=.45\textwidth,   trim =1.5cm 0.0cm 1.0cm 0cm]{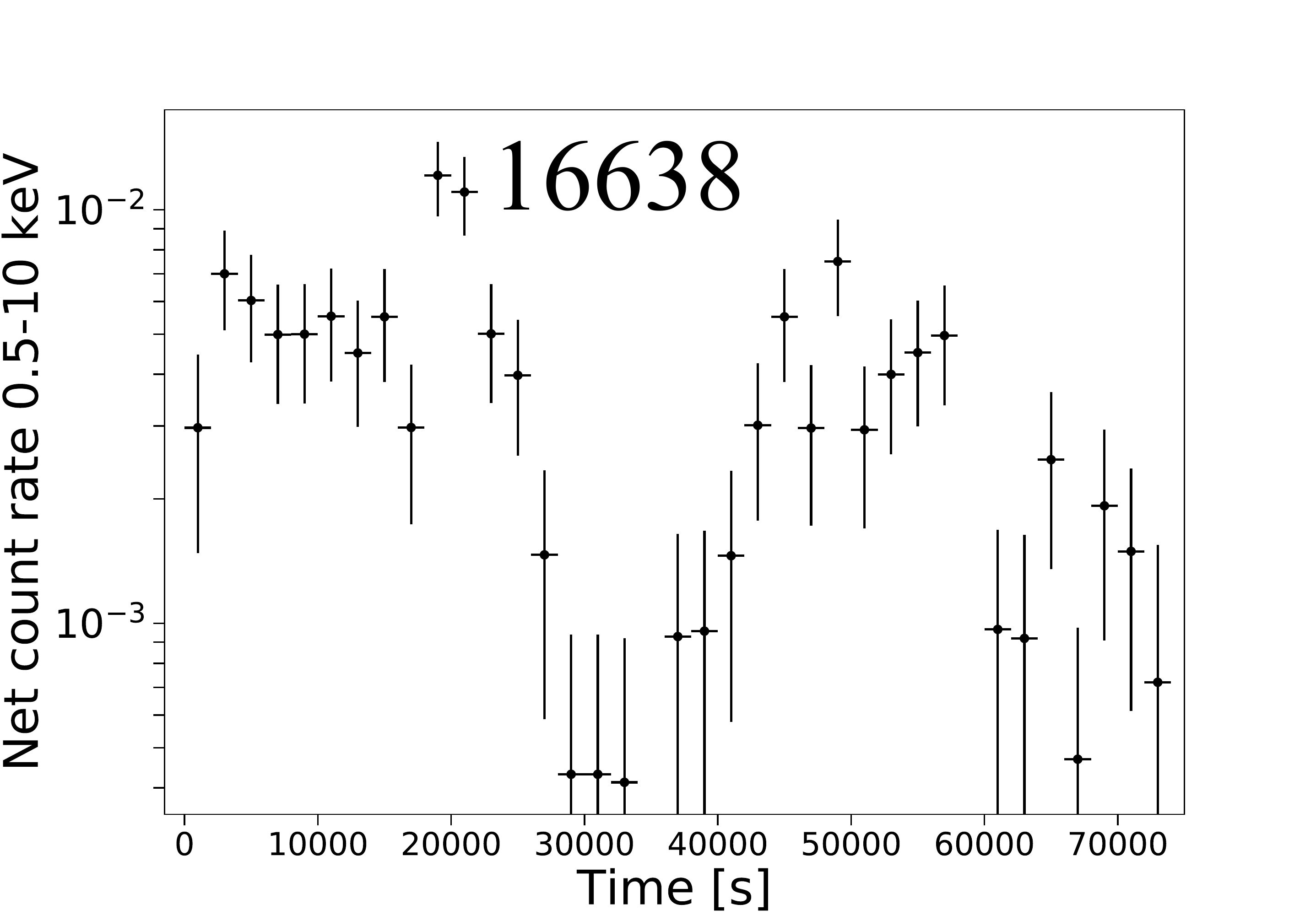}  \\
 \includegraphics[width=.45\textwidth,   trim =1.5cm 0.0cm 1.0cm 0cm]{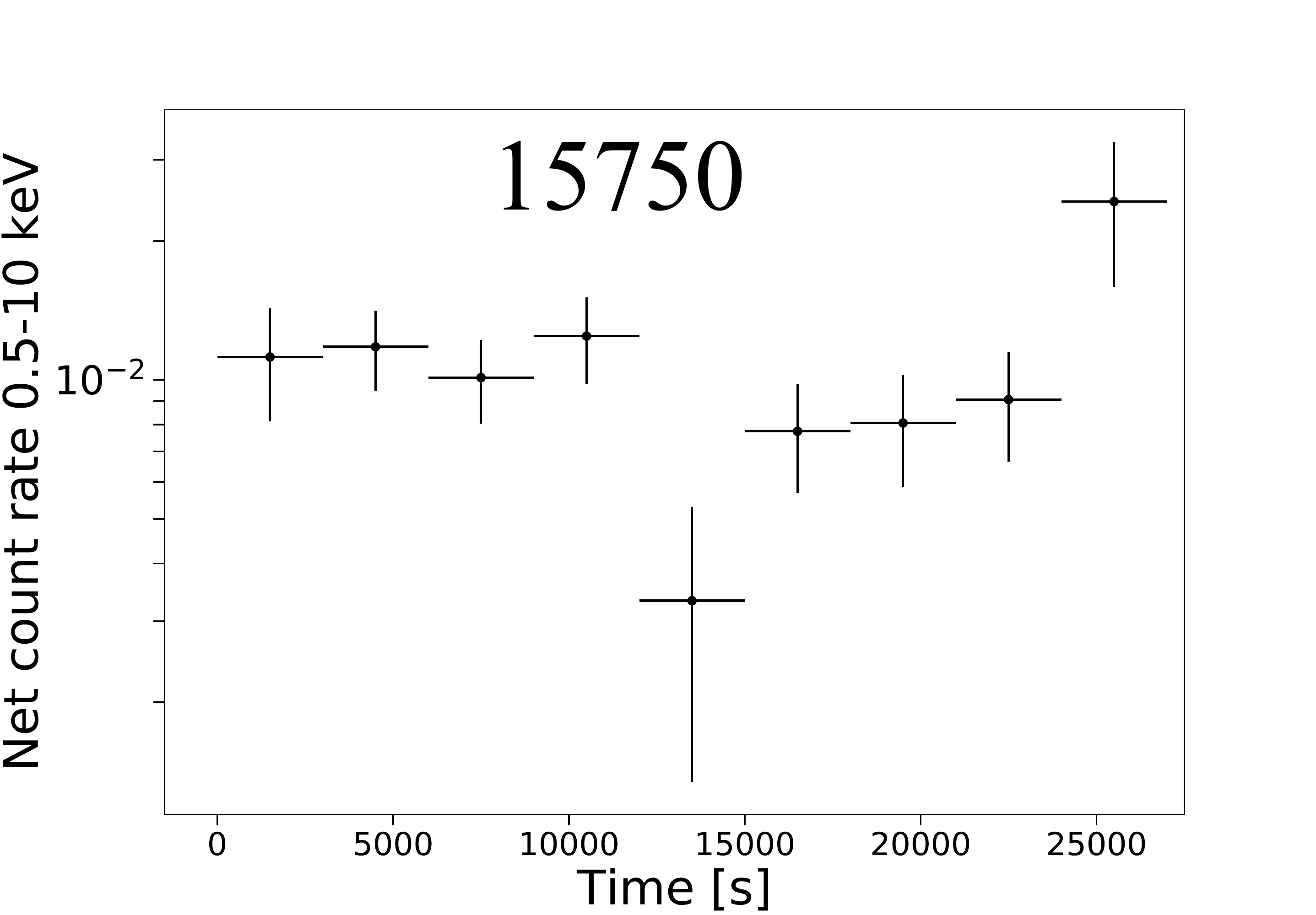}&
 \includegraphics[width=.45\textwidth,   trim =1.5cm 0.0cm 1.0cm 0cm]{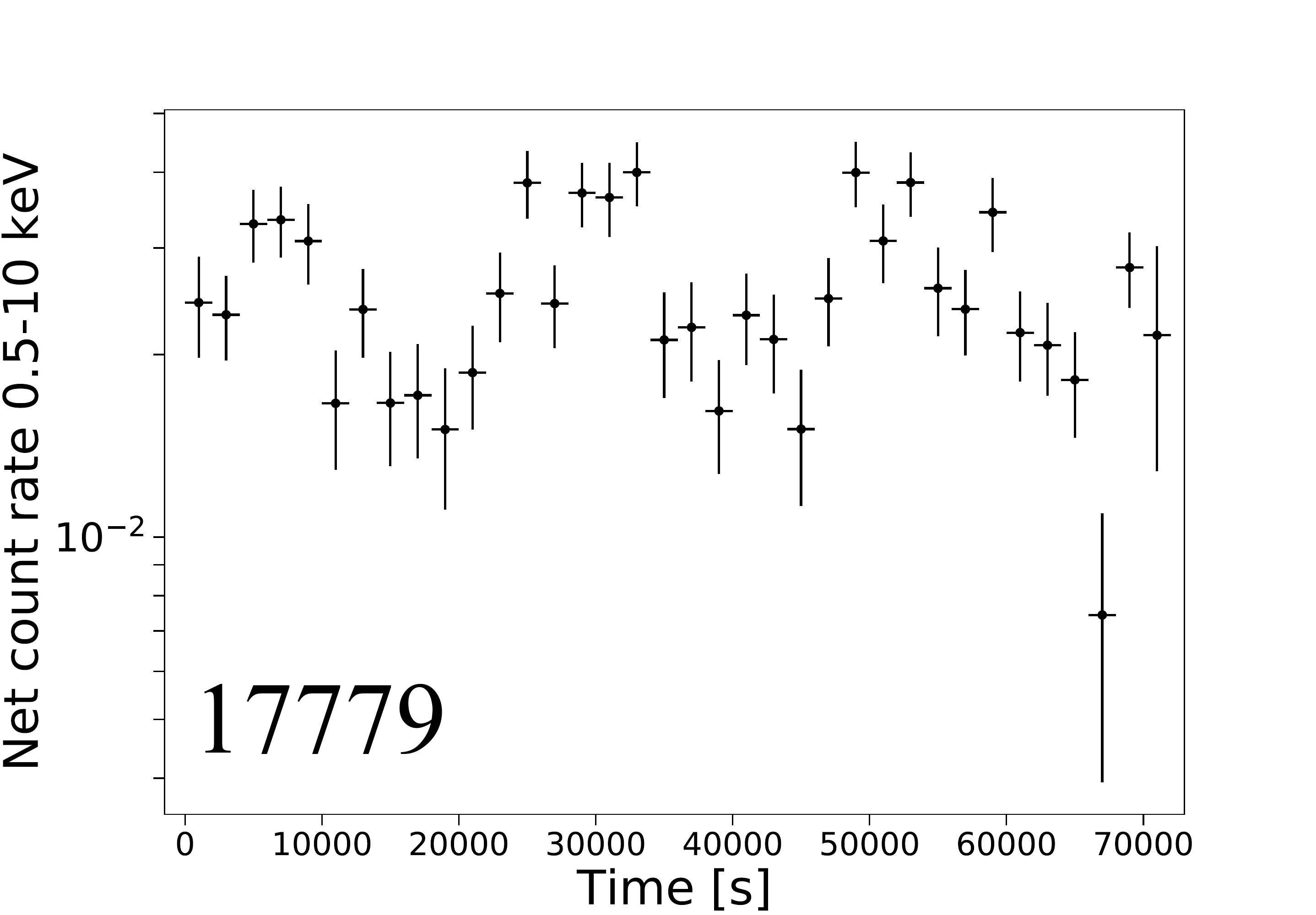}  \\
 \includegraphics[width=.45\textwidth,   trim =1.5cm 0.0cm 1.0cm 0cm]{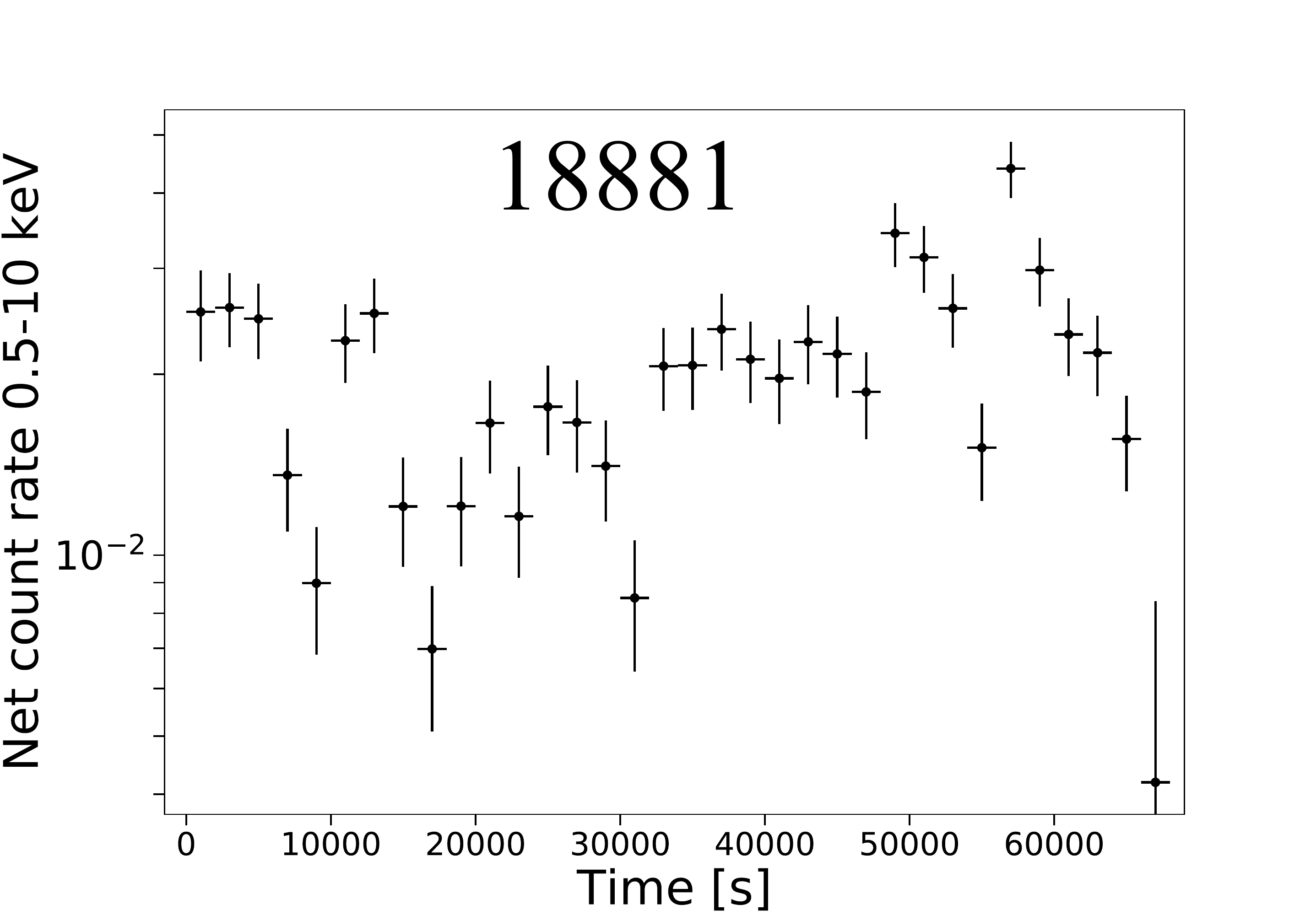}\\
  \end{tabular}
\addtocounter{figure}{-1}
 \caption{Continued}
\end{figure*}

\begin{table*}
\begin{center}
\caption{Results for the power-law model with $N_H$, $\Gamma$, and $F_X$ as free parameters (fitted in the energy range 0.5-10 keV).
All quoted errors correspond to 90\% confidence intervals. To calculate the luminosities (0.5-10 keV), we assume a distance towards EXO
1745--248 of 5.5 kpc. 
When fitting the individual groups, the $N_H$ and $\Gamma$ were tied for each observation during the fitting, but the fluxes were left free.
For each group, the X-ray fluxes and luminosities correspond to the average value of the individual fluxes.
Column 8 indicates the length of the time bins used to create the light curves shown in Figure \ref{fig:lightcurves1}.}
\begin{tabular}{ l | c | c | c | c | c|  c | c |c | c }
\hline
Date & ObsID &  N$_{H}$                                 &$\Gamma$ & $F_X$                                                    &    $L_X$      				& $\chi_{\nu}^{2}$ (dof) & Time bin & Group  \\
        &             &   ($10^{22}$ cm$^{-2}$)           &                  & ($10^{-13}$ erg cm$^{-2}$ s$^{-1}$)      &    ($10^{33}$ erg  s$^{-1}$)  & & size (ks) & \\
\hline
2003-07-13/14       & 3798 &  $1.6_{-1.2}^{+1.3}$ 		& $1.5_{-0.6}^{+0.7}$		&  $2.1^{+0.9}_{-0.4}$  	&  $0.7_{-0.1}^{+0.3}$ & 0.2 (11) &2 &--\\
2009-07-15/16       & 10059 &  $1.1_{-0.7}^{+0.8}$        & $1.0\pm0.4$  			&  $3.1\pm0.4$ 		  	&  $1.1\pm0.1$  & 1.1 (21) &2 & 2\\ 
2011-09-08	&14339&  $<1.8$ 				& $0.9_{-0.4}^{+0.8}$ 		& $0.7\pm0.2$  			& $0.3\pm0.1$  &0.7 (3) & 5 & 1	\\
2012-05-13/14	&13706& $1.4_{-0.2}^{+0.3}$ 		& $1.2\pm0.2$ 			& $7.5\pm0.4$ 	 		&    $2.7\pm0.2$    &1.04 (73)&2	 & 3	\\
2013-02-22	&14625& $1.3\pm0.4$ 			& $1.5\pm0.3$  			& $3.5_{-0.3}^{+0.4}$  	& $1.3\pm0.1$  &1.3 (40)  &2	& 2\\
2013-07-16/17	&14478& $<4.5$ 				&  $1.6_{-1.0}^{+1.4}$		 & $1.2_{-0.3}^{+0.2}$ 	&  $0.4_{-0.1}^{+0.9}$ &0.9 (4) 	& 2 &1	\\ 
2014-07-15        &14479&  $2.1\pm0.1$ 			& $1.7\pm0.1$   			& $45.5_{-1.9}^{+2.1}$	 & $16.5_{-0.7}^{+0.8}$  &	 1.1 (194) &1 &--\\
2014-07-17/18	&16638& $1.7_{-1.1}^{+1.2}$ 		& $1.5_{-0.6}^{+0.7}$ 		& $1.0_{-0.2}^{+0.4}$ 		& $0.4\pm0.1$ &0.7 (12)	&2	 &1 	\\
2014-07-20	&15750& $1.1_{-1.0}^{+1.1}$ 		 & $1.4_{-0.5}^{+0.6}$ 	& $2.8_{-0.4}^{+0.7}$ 		& $1.0_{-0.2}^{+0.3}$     &1.6 (13) &3 &2	\\
2016-07-13/14	&17779& $1.7\pm0.2$ 			& $1.6_{-0.1}^{+0.2}$  	& $8.0\pm0.5$ 			 & $2.9\pm0.2$	 &  1.04 (104)	& 2 & 3\\
2016-07-15/16	&18881&  $1.6\pm0.3$			 & $1.4\pm0.2$ 			& $5.7_{-0.3}^{+0.4}$		  &	$2.1\pm0.1$	 & 1.3 (74) & 2 & 3\\
\hline
        & Group  &             N$_{H}$                                              &$\Gamma$ & $\langle F_X \rangle$                                                    &    $\langle L_X \rangle$      \\
        &             &   ($10^{22}$ cm$^{-2}$)           &                  & ($10^{-13}$ erg cm$^{-2}$ s$^{-1}$)      &    ($10^{33}$ erg  s$^{-1}$)  \\
\hline
&1             	& $1.5\pm0.9$ &  $1.4\pm0.5$ & $1.0_{-0.3}^{+0.4}$ & $0.4_{-0.1}^{+0.2}$ & 0.7 (23)& \\
& 2                    & $1.2\pm0.3$ &  $1.3\pm0.2$ & $3.1\pm0.5$ & $1.1\pm0.2$ & 1.3 (78)&  \\
&3                    & $1.6\pm0.1$ &  $1.5\pm0.1$ & $7.0\pm0.6$ & $2.6\pm0.2$ & 1.1 (255)&  \\
\hline
Date & ObsID &  N$_{H}$                                 &$\Gamma$ & $F_X$                                                    &    $L_X$      				& W-stat (dof) & Time bin \\
        &             &   ($10^{22}$ cm$^{-2}$)           &                  & ($10^{-13}$ erg cm$^{-2}$ s$^{-1}$)      &    ($10^{33}$ erg  s$^{-1}$)  &  &size (ks) \\
\hline
2013-02-23/24	&15615& $1.0_{-0.5}^{+0.6}$ & $1.6_{-0.5}^{+0.6}$ & $0.3\pm0.1$  & $0.10\pm0.04$	 & 1.2 (89)	 &5\\
2011-02-17	&13225   & 1.4  fix & 1.4 fix  & $0.10_{-0.06}^{+0.09}$  & $0.04_{-0.02}^{+0.03}$  &   5.2 (8) &5	 \\
2011-04-29/30	&13252   & 1.4 fix  & 1.4 fix  & $0.08_{-0.04}^{+0.05}$ & $0.03_{-0.01}^{+0.02}$  & 16.6 (11) & - \\
2011-09-05         &13705  & $1.4_{-1.3}^{+1.5}$ & $1.2_{-0.8}^{+0.9}$ & $1.3_{-0.3}^{+0.6}$& $0.5_{-0.1}^{+0.2}$  & 58.8 (52) &3\\
2012-09-17/18	&14475   & 1.4 fix & 1.4 fix & $0.4\pm0.1$  & $0.10\pm0.04$ & 55.3 (33)&5 \\
2012-10-28	&14476   & 1.4 fix  & 1.4 fix & $0.30_{-0.09}^{+0.10}$ & $0.10\pm0.04$ & 44.7 (31) &5 \\
2013-02-05        &14477  & 1.4 fix & 1.4 fix & $0.20_{-0.07}^{+0.09}$ & $0.07_{-0.02}^{+0.03}$ & 21.3 (19) &5\\
\hline
\end{tabular}
\label{free}
\end{center}
\end{table*}

We found that, on top of the long-term variations, the source also displayed intense short-term variability. This is shown in the light curves obtained during ObsIDs 13706, 14625 and 14479 (Figure \ref{fig:lightcurves1}), where variations of two orders of magnitude within a few hours are observed.

\begin{figure}
\centering
  \includegraphics[width=.5\textwidth, trim=1.cm 0.0cm 0.0cm 1.5cm]{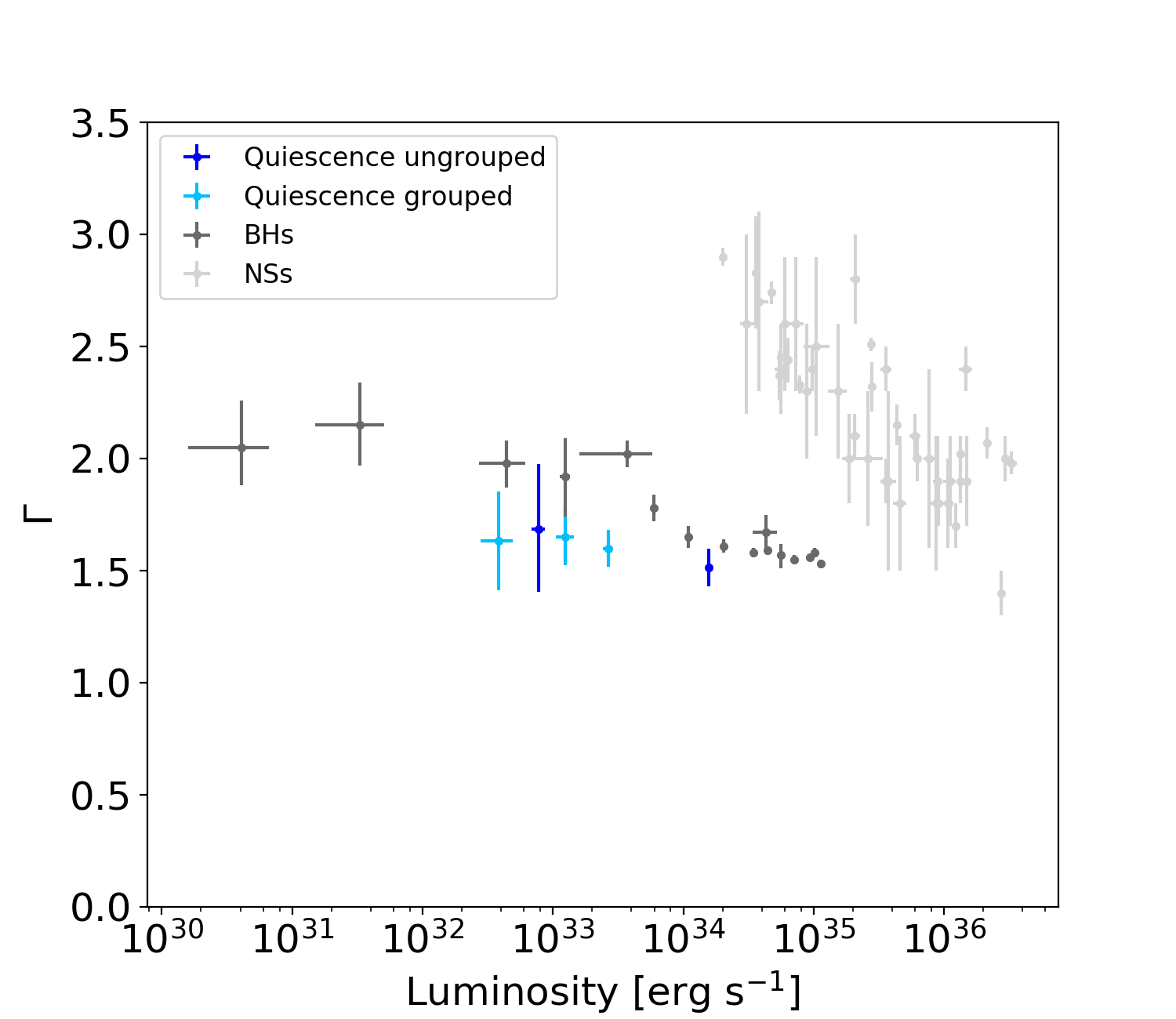}
 \caption{Similar to Figure \ref{gama_lx_free} \citep[without the very hard state points when the source was in outbursts;][]{2017parikh}, but for the case in which $N_H$ is tied between the observations during the spectral fitting.}
\label{nh_tied}
\end{figure}

Although Figure \ref{all_free} shows that $N_H$ might not be constant in quiescence (and thus not only due to interstellar absorption), it is also possible that $N_H$ might in fact remain constant. 
If the gas column is actually not changing, the inferred variability in $N_H$ would then indicate that the true intrinsic spectral shape of the source might change but be masked by the inferred $N_H$ variations. 
To investigate further if $N_H$ changes during quiescence, we fitted the spectra tying the $N_H$ between the individual observations. $\Gamma$ and $F_X$ were left free to vary. We only did this for the data for which we could use the $\chi^2$ statistics (i.e. we excluded the spectra with 1 count per bin, since they require the use of W-statistics). The results of this analysis are given in Table \ref{table_nh_tied}. 
The photon indices and fluxes are consistent within uncertainty with those presented in Table \ref{free} (where $N_H$ is free to vary). Therefore, it could be possible that $N_H$ indeed remains constant in quiescence, but leaving it free would not significantly affect the obtained photon indices and fluxes (besides making the errors bars on the indices slightly larger).
Also in this case, we then grouped observations with similar $L_X$ and $\Gamma$.
Thus, for each group we also tied the value of $\Gamma$.
We fitted the groups with the remaining individual observations (i.e. ObsIDs 3798 and 14479),
and report the results in Table \ref{table_nh_tied} (see also Figure \ref{nh_tied}).
The values of $\Gamma$ and $F_X$ for the remaining individual observations are 
practically identical to those obtained without grouping (keeping $N_H$ tied between observations) and therefore we do not show them.

\begin{table*}
\begin{center}
\caption{The fit results obtained in the 0.5-10 keV band when tying $N_H$ between all spectra, but leaving $\Gamma$ and $F_X$ as free parameters. The resulting $N_H$ was N$_{H}$  = $1.8_{-0.9}^{+0.1}$ $10^{22}$ cm$^{-2}$.
All quoted errors correspond to 90\% confidence intervals. The luminosities (0.5-10 keV) were calculated using a distance towards EXO
1745--248 of 5.5 kpc. Only those observations for which the $\chi^2$ statistics could be used were considered for the fit.  The resultant $\chi_{\nu}^{2}$ value was $1.15$ for 565 dof for the individual observations and $\chi_{\nu}^{2}=1.17$ for 571 dof for the fit in which some of the individual spectra were grouped. Groups are formed as in Table \ref{free}}. 
\begin{tabular}{ l | c | c | c | c }
\hline
Date & ObsID &                                $\Gamma$ & $F_X$            &    $L_X$      	 \\
        &             &                              & ($10^{-13}$ erg cm$^{-2}$ s$^{-1}$)      &    ($10^{33}$ erg  s$^{-1}$)  \\
\hline
2003-07-13/14       & 3798     &  $1.7\pm0.3$ &  $2.1\pm0.2$  &  $0.80\pm0.07$   	\\
2009-07-15/16       & 10059   & $1.3\pm0.2$ &  $3.4\pm0.3$ &  $1.2\pm0.1$\\ 
2011-09-08	&14339&   $1.7_{-0.5}^{+0.6}$  &  $0.9_{-0.1}^{+0.2}$ & $0.30_{-0.04}^{+0.07}$  \\
2012-05-13/14	&13706&   $1.5\pm0.1$  &  $7.8\pm0.4$ & $2.8\pm0.1$\\
2013-02-22	&14625& 	 $1.8\pm0.2$ &  $3.9\pm0.3$ & $1.4\pm0.1$\\
2013-07-16/17	&14478&  $1.8\pm0.4$ &  $1.3\pm0.2$ & $0.50\pm0.07$	\\ 
2014-07-15        &14479&  $1.5\pm0.1$ &  $43.0\pm1.4$ & $15.6\pm0.5$ \\
2014-07-17/18	&16638&	$1.6\pm0.3$ &  $1.0\pm0.1$  &  $0.40\pm0.04$\\
2014-07-20	&15750& 	 $1.7\pm0.3$ &  $3.2_{-0.3}^{+0.4}$ & $1.2\pm0.1$\\
2016-07-13/14	&17779&  $1.7\pm0.1$ &  $8.3\pm0.4$  & $3.0\pm0.1$\\
2016-07-15/16	&18881&  $1.5\pm0.1$ &  $5.9\pm0.3$ & $2.1\pm0.1$\\
\hline
&Group  &  $\Gamma$ & $\langle F_X \rangle$                             &   $\langle L_X \rangle$ \\
&           &                    & ($10^{-13}$ erg cm$^{-2}$ s$^{-1}$)      &    ($10^{33}$ erg  s$^{-1}$)  \\
\hline
&1       &   $1.6\pm0.2$ & $1.1\pm 0.3$ & $0.4\pm0.1$ \\
&2    	&   $1.6\pm0.1$ & $3.4\pm0.5$ & $1.2\pm0.2$\\
&3	&   $1.6\pm0.8$ & $7.3\pm0.6$ & $2.6\pm0.2$\\
\hline
\end{tabular}
\label{table_nh_tied}
\end{center}
\end{table*}

\subsection{The thermal component of EXO 1745--248}
\label{thermal}

Given the high quality spectrum of ObsID 14479 (during which the source was brightest), and the non fully satisfying fit obtained using an absorbed power-law model ($\chi_{\nu}^2$=1.1 for 194 dof; p--value $p=0.14$), we added a thermal component (black-body model,  \textsc{bbodyrad} in \caps{Xspec}).
When using this two-component model and assuming an emitting region with radius of 10 km at a distance of 5.5 kpc, we obtained the following parameters: $N_H = 3.8\pm0.2$ $10^{22}$ cm$^{-2}$, $\Gamma=1.5\pm0.1$, $F_X = 45\pm2.0 \times 10^{-13}$ erg cm$^{-2}$ s$^{-1}$ and a temperature 
of $0.221\pm0.007$ keV for the black-body component. This fit produced a $\chi_{\nu}^{2}=0.9$ for 193 dof, with a p--value $p=0.82$. This probability is large compared to that one obtained with the absorbed power-law model alone. 
When using this model, the power-law component contributed $\sim41\%$  of the total flux in the 0.5-10 keV range.
On the other hand, we found that the spectral fit for the second high quality spectrum (ObsID 17779) did not show substantial improvement when we added a thermal component. That spectrum is well described by a power-law model (Table \ref{free}). This does not mean that the presence of a possible thermal component is discarded in the other spectra, but that given their quality, an additional component (beyond the power-law) is not required.

\begin{figure*}
\centering
  \begin{tabular}{@{}cc@{}}
 \includegraphics[width=.45\textwidth,   trim =2.0cm 0.0cm 2.0cm 0cm]{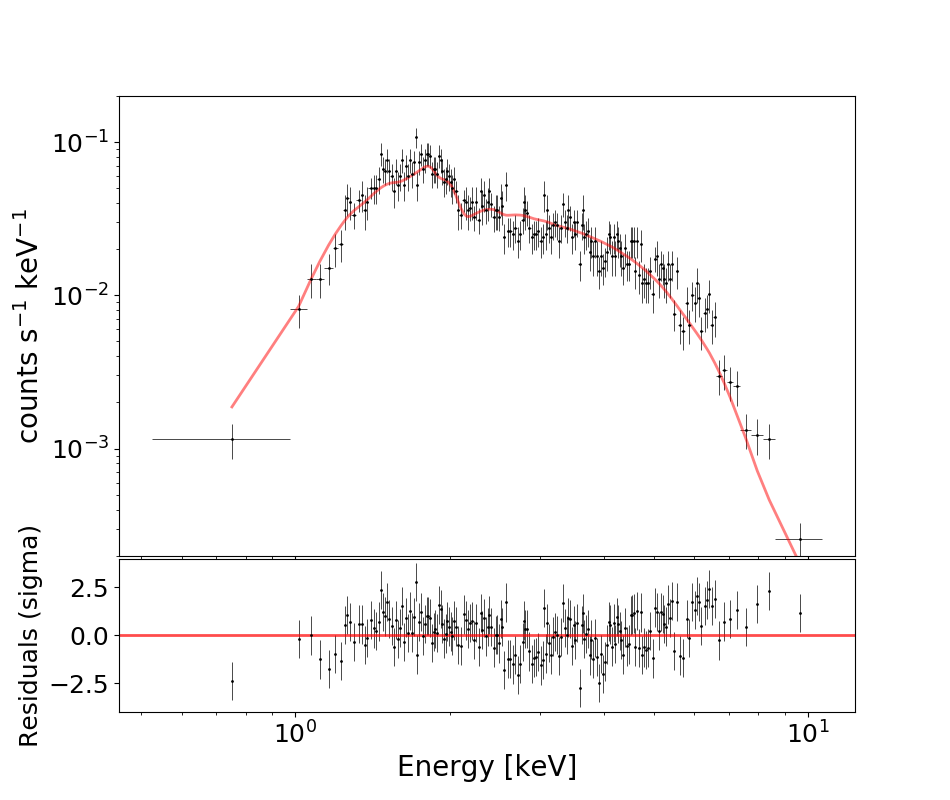}&
 \includegraphics[width=.45\textwidth,   trim =1.0cm 0.0cm 3.0cm 0cm]{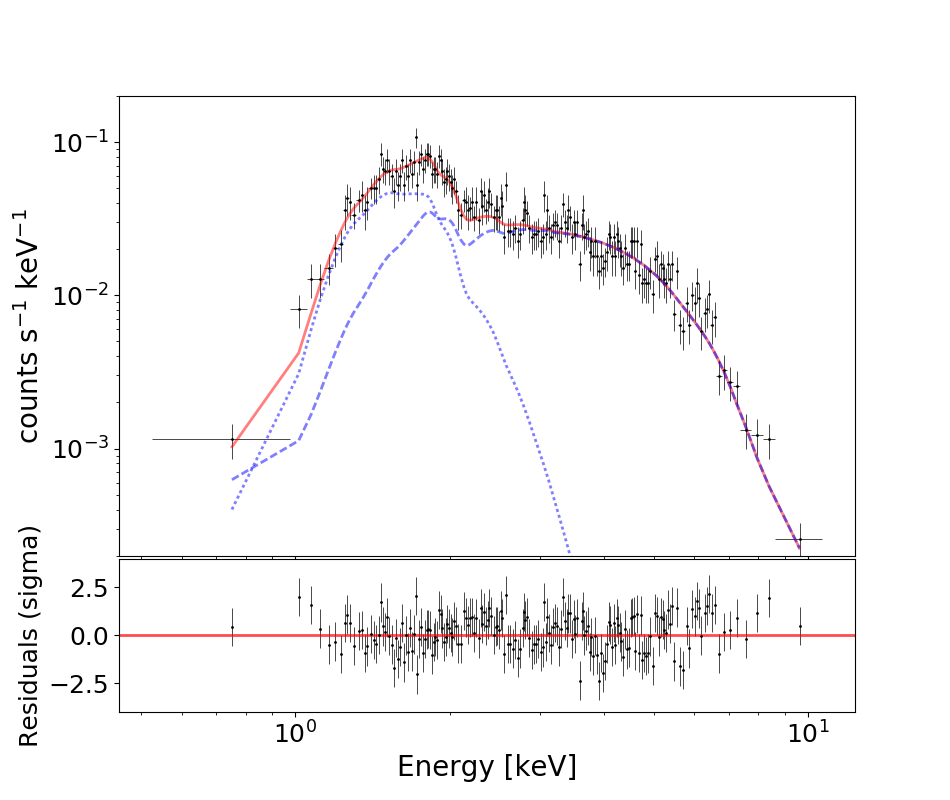}  \\
  \end{tabular}
  \caption{Comparison of the different models used to fit the spectra obtained for ObsID 14479. Left: A power-law model. Right: A model with a power-law plus a black-body component. The components are indicated with blue lines. 
The respective model is indicated with a solid red line. The residuals of each fit is indicated below in terms of sigma. }
\label{compa}
\end{figure*}

In Figure \ref{compa}, we show the spectrum (ObsID 14479) of EXO 1745--248 fitted with a power-law model (left panel) and with a two component model (right panel). 
The residuals are much smaller for the two-component model.

To see if the two component model better describes the spectrum than the absorbed power-law model alone, we have followed the method suggested by \citet{2002potra}. First, we simulated $10^5$ spectra based on the best power-law model of the original spectrum (null model). For the simulated spectra we used the same ancillary, response and background files that we used for the initial fitting. Then, we fitted a power-law model and a power-law plus black-body model to each of the simulated spectra. Using the results of each fit, we computed the F-statistics for each spectrum. We built a probability distribution with these results. Thus we calculated the probability of obtaining the same F-value that we obtained with the initial power-law fit and the fit to the spectrum using a power-law plus black-body component. We found a probability $p=0$, because the initial F-value is outside of the probability distribution. 
This means that the probability that the new component is only fitting a random fluctuation is $<10^{-5}$.
The obtained probability gives us confidence that the spectrum is better described by adding another component than by a single power-law model.

In the two components case, where we leave the radius of the thermal emitter free to vary, the obtained parameter values are:  $N_H = 2.6\pm0.2$ $10^{22}$ cm$^{-2}$, $\Gamma=1.0\pm0.1$, $F_X = 38\pm1.5 \times 10^{-13}$ erg cm$^{-2}$ s$^{-1}$, a temperature of $0.34_{-0.03}^{+0.02}$ keV and a radius of $2_{-1.0}^{+1.7}$ km for the thermal source. The $\chi^2$ fitting gives a p--value $p=0.96$. Again, we compared this model to the single absorbed power-law model. 
We followed the same steps as before to produce
an empirical probability distribution for F. We obtained a probability $p\sim5\times10^{-3}$ that the improvement in the $\chi^2$ statistic would be due only to chance. This suggests that the spectral fit likely requires a second component, which could be a black-body one.
In this two components fit when the radius of the thermal emitter was allowed to vary, the power-law component contributed $\sim64\%$ of the total 0.5-10 keV flux. Given the small radius obtained for the region of thermal emission, 
it is unlikely that it originates in an accretion disk.

We note that for the model in which we fix the radius of the black-body component, the obtained $N_H$ is very high ($\sim3.8\times10^{22}$ cm$^{-2}$). This is even higher than what has been observed when the source was in outburst (Figure \ref{all_free}) and higher than the average value of $N_H$ for the cluster itself \citep[$2.6\times10^{22}$ cm$^{-2}$;][when Wilm abundances are used]{2015bara}. But when leaving the radius as a free parameter, the $N_H$ decreases to $2.6\times10^{22}$ cm$^{-2}$ which is more in line with the other observations and with the average of the cluster, so this model might be more applicable. However, it is still unclear why the $N_H$ in this model is still significantly higher than observed for the other quiescent observations. This likely indicates correlated behavior between $N_H$ and the other model parameters.  
However, it is important to note that despite adding a thermal component (which substantially improved the fit), the model shows residuals 
which suggest that a more complex model could improve even more the fit of the data. But, in particular it is clear that a single power-law spectrum does not describe properly the spectrum of ObsID 14479. Instead, an additional component should be added. This in turn could explain the obtained high value of $N_H$. 

\section{Discussion}

We report on 18 \textit{Chandra} observations (spanning from 2003 to 2016) performed on the GC Terzan 5 to study the quiescent properties of the transient NS LMXB EXO 1745--248. We found that the source is extremely variable: its quiescent luminosity (assuming a distance of 5.5 kpc; 0.5-10 keV) varies between $\sim3 \times 10^{31}$ erg s$^{-1}$ to $\sim1.7 \times 10^{34}$ erg s$^{-1}$ on timescales of days to years. Similarly to what was found previously by \citet[][]{2005wijnands} and \citet{2012dege}, we obtained quiescent spectra that are very hard, with photon indices of $\sim$1.4 (when the spectra are fitted with a single power-law model). Intriguingly, the spectral shape does not significantly change over this large luminosity range (although the column density is potentially correlated with the luminosity\footnote{As already suggested by \citet[][]{2012dege}, this might be the reason why they found that the X-ray colors of the source during the 2003 observation (ObsID 3798) were softer than those obtained during the 2009 observation (ObsID 10059) despite that we have found that the photon indices during both observations are consistent with each other.}; Figure \ref{all_free}), except for the brightest observation during which a soft component was added to obtain a better fit (Section \ref{thermal}). Note that a soft thermal component is frequently observed in quiescent NS LMXBs \citep[see e.g. X3 in Terzan 5 and GRS 1747--312 in Terzan 6; ][]{2015deg_c,2018_vats} when spectra of similar S/N are available.

\citet[][]{2005wijnands} already reported on strong variability of the quiescent emission of EXO 1745--248. Also
\citet[][]{2012dege} found that the source varied by a factor of $\sim$3 within hours, but by an order of magnitude between different observations (separated by years). Here we demonstrate that the variability of the source is even more extreme: its average luminosity varies by up to two orders of magnitude within $\sim$4 hours (e.g., during observations with ObsID 13706, 14625 and 14479; Figure \ref{fig:lightcurves1}) to nearly three orders of magnitude on the timescale of years. Interestingly, the source was brightest during the observation performed on July 15th, 2014 (ObsID 14479), with an X-ray luminosity of $\sim1.7 \times 10^{34}$ erg s$^{-1}$. But only a few days later (on July 17th/18th, 2014; ObsID 16638), the source had decreased again down to $\sim 4\times 10^{32}$ erg s$^{-1}$. However, during both observations the source was very variable as well (see Figure \ref{fig:lightcurves1}), 
with a lowest count rate of only a few times $10^{-4}$ counts s$^{-1}$ (ObsID 16638) and with a highest one of $\sim3\times 10^{-1}$ (ObsID 14479). This indicates count rate fluctuations in only three days of nearly three orders of magnitude. 
The mechanism behind this strong variability of the source is currently not clear.

\citet[][]{2012dege} discussed in detail the potential origin of the quiescent emission of EXO 1745--248 and concluded that the emission likely was due to very low level accretion down to the NS. However, recent observational insight \citep{2010cacket,2014cha,2015dangelo,2015wij} suggests that when low level accretion onto a NS occurs, the resulting X-ray spectrum would have a soft thermal component (released when the matter hits the NS surface) as well as a hard, non-thermal one (due to Bremsstrahlung in the boundary layer). This is quite contrary to what we observe for EXO 1745--248. Only for the brightest observation a soft component substantially improved the fit. The soft component could still be present at lower luminosities. However, due to the limited quality of the spectra an extra component is not statistically required. So, if indeed EXO 1745--248 is accreting at very low levels in its quiescent state, then (at least) below $10^{34}$ erg s$^{-1}$ some other emission mechanism might be at work compared to the other quiescent NS LMXBs that are still accreting.

In this context, it is interesting to compare our source with the transient IGR J18245--2452 in the GC M28 \citep{2013papitto}. This source is one of the so-called transitional millisecond pulsars (tMSPs), which are systems that transition between a radio millisecond pulsar phase and a phase in which they are accreting. IGR J18245--2452  is the only tMSP for which a bright ($\sim10^{37}$ erg s$^{-1}$) outburst has been observed \citep{2013papitto,2014linares,2017falco}. During the outburst, the source showed several remarkable properties. Its spectra were always very hard \citep{2014linares,2017parikh} and it showed very strong variability
(\citeauthor{2013papitto} \citeyear{2013papitto}; \citeauthor{2017falco} \citeyear{2017falco}; \citeauthor{2017wijnandsb} 
 \citeyear{2017wijnandsb}b). In addition, also its quiescent spectra are very hard and strongly variable \citep{2014linares}. 

This looks very similar to what we observed for EXO 1745-248. Both sources exhibit a highly unusual very hard state at relatively high luminosities \citep{2017parikh}. In particular, EXO 1745-248 was also very hard during (part of) its 2015 outburst \citep[][]{2016teta} and strongly variable (\citeauthor{2017wijnandsb} 2017b). Moreover, strong variability was also seen during its 2000 outburst (Altamirano et al. in preparation). Finally, also resembling IGR J18245-2452, the quiescent spectrum of EXO 1745--248 is also very hard and extremely variable (see Section \ref{analysis}). \citet{2015ferraro} identified the optical companion of EXO 1745--248 as a main sequence turn-off star which is experiencing its first envelope expansion. These authors suggested that it is possible that the system is in an evolutionary state previous to the swinging phase shown by tMSPs.

The extreme properties of IGR J18245--2452 have been attributed to the effect of the NS magnetic field in this system \citep[see, e.g., the discussion in][]{2014ferri}. This might suggest that something similar is possible for EXO 1745--248, and could explain why during our brightest quiescent observation (ObsID 14479) the source showed a thermal contribution potentially arising from only a small emitting region (see Section \ref{thermal}): the presence of a dynamically important magnetic field would potentially channel the accreted material only to the magnetic poles of the NS and not to the whole surface.
This scenario would result in pulsations in the data of EXO 1745--248, but the time resolution of the ACIS detectors does not allow to search for them. However, detailed pulsation searches have been performed using {\it XMM-Newton} observations obtained during the 2015 outburst of the source and none were found \citep[upper limit in amplitude of 2\%;][]{2017matra}. Similarly, no pulsations were detected during the 2000 outburst (Altamirano et al. in preparation) using the Proportional Counter Array on board of the Rossi X-ray Timing Explorer, which had a higher sensitivity to pulsations than \textit{XMM-Newton}. 
Although weak pulsations might still be present for EXO 1745--248, it is unlikely that this system is, in this aspect, similar to IGR J18245--2452 (see also the discussion in \citeauthor{2017wijnandsb} 2017b). Moreover, geometrical effects could prevent the detection of pulsations.

Another difference between IGR J18245--2452 and EXO 1745--248 is the fact that the observed variability in IGR J18245--2452 was quite different during outburst (i.e., even more extreme) than observed for EXO 1745--248, 
potentially indicating different types of accretion flows (see \citeauthor{2017wijnandsb} 2017b).
On the other hand, during quiescence, the short-term variability properties of EXO 1745--248 (Figure \ref{fig:lightcurves1}) are more extreme (and less regular) than what has been seen for IGR J18245--2452 \citep{2014linares}. 
Finally, also the outflow properties are quite different: IGR J18245--2452 is a very strong (for a NS transient) radio source\footnote{During outburst and in quiescence, IGR J18245--2452 is a strong radio emitter, but it is only in quiescence that the radio emission is pulsed.} during outburst \citep[L $\sim 10^{37}$ erg s$^{-1}$,][]{2013pavan,2013papitto}, but EXO 1745--248 is actually the faintest NS radio source known at similar luminosities \citep[when compared to NS transients in the hard or very hard state;][]{2016teta}. This again suggests quite different accretion properties between both systems. So despite that the systems are both very hard over a very large luminosity range and show very strong variability, the details of their behavior differ.  Thus, it is unclear 
whether we see the same accretion process active in both systems with only small changes between sources (such as the strength or configuration of the magnetic fields of the NSs) or that two, likely unrelated mechanisms are at work.

For EXO 1745--248 we have observed X-ray luminosities ranging up to $\sim1.7 \times 10^{34}$ erg s$^{-1}$ and the source is always very hard over the full luminosity range we explored. \citet{2017parikh} report on \textit{Swift} outburst data of this source \citep[see also][]{2016teta} and found that around $\sim10^{36-37}$ erg s$^{-1}$, EXO 1745--248 had also very hard spectra (it was one of the only three NS systems identified by these authors to have this previously unrecognized very hard state). 
However, their data only goes down to a luminosity of $\sim10^{36}$ erg s$^{-1}$ (see also Figure \ref{gama_lx_free}), with a potential minor softening at the lowest observed luminosity. 
The spectral behavior of the source is unknown in the so far unexplored luminosity range of $10^{34}$--$10^{36}$ erg s$^{-1}$.
It might be possible that the source always remained hard and, in this case, it would behave quite differently from what is observed for the majority of the NS systems \citep[see Figure \ref{gama_lx_free} and Figure 1 in][]{2015wij}. 

The softening of the other NS systems is thought to be due to the NS surface starting to dominate their X-ray spectra \citep{2015wij}. If for EXO 1745--248 
the contribution from the NS surface is lower (in our preferred two component model the thermal component only contributes $\sim36\%$ to the 0.5-10 keV flux instead of the usual 50\%), then the source indeed could remain hard over a very large luminosity range (from a few times $10^{31}$ erg s$^{-1}$ to $\sim10^{37}$ erg s$^{-1}$). In this case it is interesting to compare our source once again with IGR J18245--2452 because \citet[][see also \citeauthor{2017parikh}, \citeyear{2017parikh}]{2014linares} found that the latter indeed remained hard down to $\sim 10^{35}$ erg s$^{-1}$, but softened (for unclear reasons) between $10^{34}$ and $10^{35}$ erg s$^{-1}$. 
However, it became hard again at lower luminosities. 
Observations during a future outburst of EXO 1745--248 could test if a similar behavior is observed to that of IGR J18245--2452. 

Alternatively, EXO 1745--248 might follow the same track as observed for the other NS systems (see Figure \ref{gama_lx_free}), but if true, at around $10^{34}$ erg s$^{-1}$ it suddenly would have to become hard again (i.e., the NS would suddenly not be visible again; or only with a relatively small contribution). This would be roughly similar to the behavior of the accreting millisecond pulsar SAX J1808.4-3658 \citep{1998wijnands} which has a similar hard quiescent spectra \citep{2002campana,2007heinke,2009heinkeb}, but above $10^{34}$ erg s$^{-1}$ it follows the track of the other NS systems \citep[see Figure 1 of][]{2017parikh}. However, the quiescent luminosity of SAX J1808.4-3658 is very low, $\sim8 \times 10^{31}$ erg s$^{-1}$ and the transition 
from the soft spectra to the hard spectra below $10^{34}$ erg s$^{-1}$ could be more gradual than what is required for EXO 1745--248. More detailed spectral observations are needed for all three sources (EXO 1745--248, IGR J18245--2452 and SAX J1808.4-3658) below $10^{36}$ erg s$^{-1}$ to be able to make more conclusive statements about the similarities and differences between them, which would highlight potential differences in NSs and/or accretion properties.

\section*{Acknowledgments}
We thank the anonymous referee for his/her comments on the manuscript. 
LERS acknowledges support from NOVA and a CONACyT (Mexico) fellowship. 
RW acknowledges support from a NWO top grant, Module 1. 
ND is supported by a Vidi grant awarded by the NWO. 
YC is supported by the European Union Horizon 2020 research and innovation programme under the Marie Sklodowska-Curie Global Fellowship grant agreement No 703916.
DA acknowledges support from the Royal Society.
CH was supported by an NSERC Discovery Grant and Discovery Accelerator Supplement. 
The NASA ADS abstract service was used to access scientific publications and for getting the references used in this paper. Support for this work was provided by the NASA through $Chandra$ Award Number GO6-17031B issued by the \textit{Chandra X-ray Observatory Center}, which is operated by the Smithsonian Astrophysical Observatory for and on behalf of the NASA under contract NAS8-03060.
\bibliographystyle{mn2e}
\bibliography{ref_exo_sub}

\clearpage

\end{document}